\newcommand{\eqref}[1]{Eq.\ (\ref{#1})}

\newcommand{\figref}[1]{Fig.\ \ref{#1}}
\newcommand{\beq}{\begin{equation}}
\newcommand{\enq}{\end{equation}}
\newcommand{\beqarr}{\begin{eqnarray}}
\newcommand{\enqarr}{\end{eqnarray}}
\documentclass[12pt]{article}
\usepackage{graphicx}
\begin{document}
\title{A Landau Primer for Ferroelectrics}         
\author{P. Chandra and P.B. Littlewood}
\date{\today}
\maketitle
\section{Introduction} 

Based solely on symmetry considerations, Landau theory 
can provide a reliable description of a system's equilibrium behavior
near a phase transition. This phenemenological 
approach relates measurable quantities to one another using a minimum
set of input parameters that can be determined either by
comparison to experiment or from first-principles approaches.  Landau
theory (LT) can therefore serve as a conceptual bridge between
microscopic models and observed macroscopic phenomena.
Because it assumes spatial averaging of all local fluctuations,
LT is particularly well-suited to systems with long-range interactions
such as superconductors and ferroelectrics.  In this chapter we hope
to give the curious reader a self-contained contemporary presentation of
the application of Landau theory to ferrolectrics, including references
to more extensive treatments for those who wish more 
detail \cite{Landau59,Toledano87,Fatuzzo67,Lines77,Jona93,Strukov98}.
We begin by developing LT for homogenous bulk ferroelectrics, and then
consider the finite-size (thin film) case \cite{Scott00} within the
related Landau-Ginzburg (LG) approach.  Next we discuss the treatment of
inhomonogeneity within this framework.  We end with a number of
open questions for future pursuits.

Let us begin by stating in general terms what Landau theory is and then
subsequently what it is not.  In a nutshell, 
Landau theory is a symmetry-based analysis of equilibrium
behavior near a phase transition. In his classic 1937 papers, 
Landau \cite{Landau37} 
notes that a system cannot change smoothly between two phases of different
symmetry; the continuous path that a system
can take between liquid and gaseous states around a first-order critical
point is only possible because there is no symmetry change involved.
Furthermore, because the thermodynamic states of two phases 
that are symmetrically distinct 
must be the same at their shared transition line, 
the symmetry of one phase must be higher than that of the other.  Landau 
then characterizes the transition in terms of an order parameter, a 
physical entity  
that is zero in the high-symmetry (disordered) phase, and changes continuously
to a finite value once the symmetry is 
lowered; for the case of
the paraelectric-ferroelectric transition \cite{Mueller40,Devonshire49}, 
this order parameter is
the polarization $P$.  
The free energy, ${\cal F}$, in the vicinity of the transition is then expanded
as a power series of the order parameter (${\cal F}(P)$)  
where only symmetry-compatible
terms are retained. The state of the system 
is then found
by minimizing
this free energy ${\cal F}(P)$ with respect to $P$
to obtain $P_0$,
and specific thermodynamic functions are subsequently computed
by differentiating ${\cal F}$ accordingly.
As an aside, we note that if the order parameter is multicomponent, then
the Landau free energy is constructed from all scalar terms consistent
with the system's symmetries that are powers and scalar products
of the order parameter components.  Here, for simplicity, we'll consider
a scalar order parameter ($P$) unless otherwise specified.

%************************************************************************
\begin{figure}
\begin{center}
\includegraphics[height=2.5in]{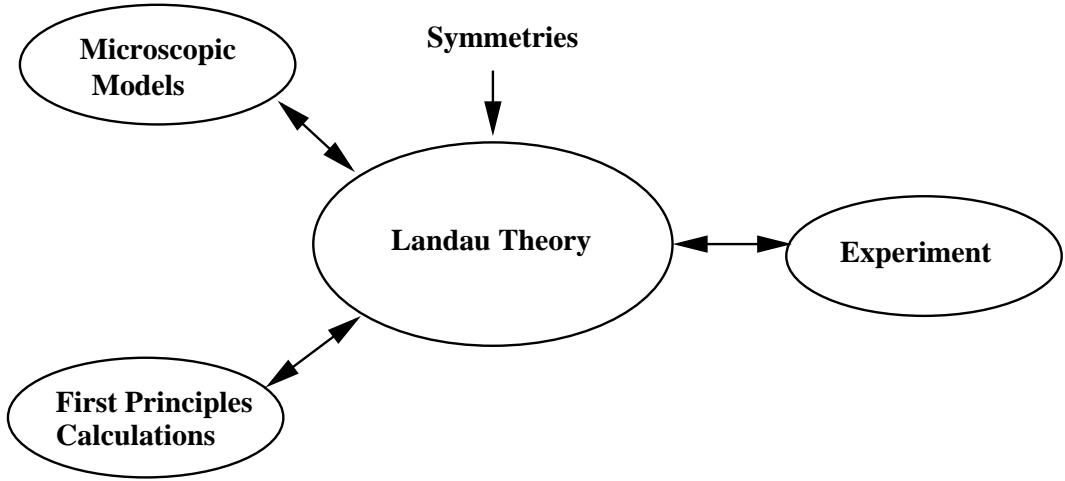}
\end{center}
\caption{\label{fig:landau} Schematic of relationship of Landau theory and
first principles calculations, microscopic models and experiment in the 
vicinity of a phase transition}
\end{figure}
%***************************************************************************

The appeal of the Landau approach
is that it is a straightforward phenemenology for linking 
measurable thermodynamic quantities
in the vicinity of a phase transition.  However it is only as good as its
input parameters, the coefficients of the series 
expansion terms in ${\cal F}(P)$, 
which can be determined from experiment or from first principles 
calculations.  
We emphasize that Landau theory is strictly a macroscopic approach and thus it
cannot describe any microscopic physics (e.g. atomic displacements etc.)
associated with the phase transition.  
In its simplest rendition, the order parameter (e.g. $P$) 
is considered to be completely uniform.  

%**************************************************************************
\begin{table}
\begin{center}

\begin{tabular}{|c|c|}
\hline 
& \\
{\bf Phenomenology}&
{\bf Ferroelectric}\\
& {\bf (near $T_c$)}\\
& \\
\hline
\hline
& \\
Landau - Devonshire Theory & Poled Bulk System \\
(Uniform Polarization)&\\
& \\
\hline
& \\
Landau-Ginzburg Theory & Bulk System\\
(Polarization with Spatial Gradient)&\\
& \\
\hline
&  \\
Landau-Ginzburg Theory & Film\\
with Boundary Conditions & \\
& \\
\hline

\end{tabular}
\end{center}
\vskip0.2in
\caption{Phenomenologies and Relevant Ferroelectric Systems (close to $T_c$)}
\label{symmetries}
\end{table}

%***********************************************************************

A central ansatz of the Landau approach is that the
free energy can be represented as a series expansion of the order
parameter in the vicinity of the transition.  At first glance,
it may be surprising that singular behavior associated with a transition
can emerge from such a regular expansion.  This occurs because the 
value of the order parameter that minimizes the free energy (e.g. $P_0$)
is itself a singular function of the expansion coefficients which are 
temperature-dependent. However, more generally we expect that this
power-law form of ${\cal F}$ may not be valid very close to the transition.
Furthermore Landau theory is based on the premise that local fluctuations
in the order parameter are small, whereas we expect that this will
not be the case in the immediate vicinity of the transition.

So when does Landau theory break down?  Levanyuk and Ginzburg have 
developed \cite{Levanyuk59,Ginzburg60}
a criterion which uses Landau theory to estimate its own demise.  More
specifically, they argue that for LT to be valid, fluctuations of the order 
parameter must be small compared to the order parameter itself when both
are coarse-grained over a volume determined by the  correlation length..  
We shall present
a detailed version of their argument once we introduce a specific Landau
free energy.  Qualitatively their criterion suggests that LT works well
when the coordination number is high, which can be achieved either by large
dimensionality or by long-range interactions.  Since significant 
dipolar interactions are present in ferroelectrics, does this 
mean that Landau theory
is always valid in these systems?  It is
certainly very useful in ferroelectrics but we emphasize that the 
angular dependence of the dipolar forces leads to non-Landau behavior very
close to the transition, though this region is usually practically
unobservable (we will be more specific later).  
Indeed historically the theoretical study of a 
simplified version of this problem, the uniaxial dipolar model,
yielded the first predictions of logarithmic corrections to Landau exponents
\cite{Larkin68}.
Detailed measurements confirmed these quantitative results in appropriate
experimental systems \cite{Ahlers75}, and this set of findings served as
an important
benchmark in the development of the renormalization
group method.  

Many readers may have encountered the phenomenology of Landau theory in 
the context of magnetism, and thus it is worth  
contrasting its application there and in ferroelectrics (and we encourage
the interested reader to pursue more on this subject in the chapter by 
N. Spaldin).
First and perhaps foremost, the couplings in a ferromagnet are
predominantly short-range, since interactions between magnetic dipoles
are essentially negligible.  More specifically we note that
the interaction energy between two magnetic dipoles, each of magnitude one
Bohr magneton, at a distance $r$ is 
\begin{equation}
 F_M = \frac{\mu_o \mu_B}{4 \pi r^3} \approx \frac{\alpha^2}{4\pi}\left ( \frac{a_B}{r} \right ) ^3 
\end{equation}
where 
$a_B $= 0.05 nm and $\alpha \approx \frac{1}{137}$, and we are working in atomic units. 
By contrast, for an electrical dipole of magnitude $p = e \Delta a_B$,
the dipolar interaction force is
\begin{equation}
F_D = \frac{p^2}{8 \pi \epsilon_o r^3} \approx \frac {\Delta^2}{4 \pi} \left ( \frac{a_B}{r} \right)^3
\end{equation}
where the parameter $\Delta \approx 1$ is determined by 
examining atomic displacements 
and transverse effective charges \cite{Born54}. 
The ratio of the ferroelectric to the ferromagnetic dipolar forces 
is then of order $\frac{\Delta^2}{\alpha^2} \approx (137)^2$, so clearly 
long-range interactions
are more significant in ferroelectrics than in generic magnetic
systems.
Thus LT is more reliable for a broader temperature region in
these electric dipolar systems than in their magnetic counterparts, though
it can provide guiding
information in both cases.  In conventional ferromagnets
magnetic dipolar interactions play a role in the determination of domain
structures, whereas in ferroelectrics the analogous dipolar couplings
are important at essentially all length-scales.

Screening
represents a second qualitative distinction between ferroelectric and
ferromagnetic systems. This phenomenon is due to the presence of free charges
that effectively reduce local fields; it does not have a magnetic
analogue due to the absence of free magnetic monopoles.
Indeed this simple fact about screening has been used to explain
why magnetism was known in ancient times but ferroelectricity is a relatively
modern discovery \cite{Lines77}.  
The surface of a ferroelectric crystal attracts free charge
from its environment to neutralize the effective surface charge that results
from the polarization discontinuity; this partially accounts for the historical
name ``pyroelectric'' to describe transient fields that were observed
upon changes in temperature.  Screening effects are particularly pronounced
in ultrathin ferroelectric films, \cite{Batra72,Mehta73,Wurfel73,Ghosez00,
Junquera03,Dawber03} 
where internal electric fields
generated by the ferroelectric dipoles are large enough to produce both
electronic and ionic currents.  These processes contribute to
fatigue, to ``imprint'' and to the memory of domain patterns stored over
time \cite{Scott00}.  A detailed understanding of these screening processes, particularly
near metal-ferroelectric interfaces in thin-film devices, requires 
knowledge of metal-induced gap states and charge compensation processes.
Since these are local effects, they can be captured within
a Landau-Ginzburg framework and this physics will be discussed later in the chapter.

The third major difference between LT treatments of ferroelectricity
and ferromagnetism arises from the strong coupling between the polarization
and the lattice, a feature which is usually not very important in the
magnetic case.  Ferroelectric dipoles are generated by the displacement
of atoms, and thus on a local scale there is strong spatial anistropy.  Many
magnets retain continuous spin symmetry (e.g. $xy$ or Heisenberg), and
thus have low barriers associated with the reorientation of the magnetization.
By contrast in ferroelectrics domain walls, discontinuities in the polarization
direction, are roughly a few lattice constants in contrast to the mesoscopic
length-scale associated with their magnetic counterparts.
In general the development of a macroscopic polarization in a ferroelectric
will be accompanied by a macroscopic strain, and thus ferroelectricity
and ferroelasticity are closely related phenomena.
Indeed if this strain is suppressed, the transition temperature and even
its order can be modified.  Elastic degrees of freedom thus must
be represented in a LT for most ferroelectrics.  

Let us now turn to a general outline of this chapter.
We will begin with a discussion
of the homogenous Landau theory for bulk
ferroelectrics with spatially-uniform polarizations, 
reviewing first- and
second-order transitions and the dielectric response.
Next we allow the polarization to vary with position within
a bulk sample,
and discuss Landau-Ginzburg theory and correlation functions.
Surface boundary conditions, corresponding to the situation for thin
films, are then added and the resulting polarization profile is discussed.
Here screening effects, depolarization fields and strain will be
important.
The enigmatic extrapolation length, borrowed from the literature
on surface superconductivity \cite{Tilley84}, will be described with remarks 
about its underlying physical meaning.  Then we'll switch to a treatment
of inhomogenous systems in the bulk, where strain effects and domain 
formation become very important.  We end with a number of open questions
for further exploration.  For example, is LT appropriate
when both long-range strain and Coulomb interactions are present?  
How can we begin to treat the ever-important question of dynamics of
domain motion within this framework? Landau-Ginzburg theory is certainly
only useful when its constructing assumptions and its input parameters
are appropriate; if so, it is a fairly powerful approach, based almost
solely on symmetry, to yield important information about measurable
quantities in the vicinity of a phase transition.  In this chapter we
emphasize the underlying philosophy and methodology of this
phenomenology, pointing the interested reader to
detailed discussions of successful applications to particular materials
including $SrTiO_3$ \cite{Haeni04},
$BaTiO_3$ \cite{Li05a} and $PbZr_{1-x} Ti_x O_3$ \cite{Li05b}; 
towards this more pragmatic end,
an Appendix of Landau parameter sets for specific common ferroelectric 
systems is included for easy reference.

\section{Landau-Devonshire Theory}

\subsection{General Phenomenology}

To our knowledge, Landau's symmetry-based treatment
of phase transitions \cite{Landau59,Landau37} was first applied to
the case of ferroelectrics by Devonshire \cite{Devonshire49},
and in this section we will discuss this Landau-Devonshire
theory that is appropriate for bulk systems with spatially
uniform polarization.
In general, the thermodynamic state of any system in equilibrium can be
completely specified by the values of specific variables; for bulk
ferroelectrics these include the temperature ($T$), the polarization ($P$), 
the electric field ($E$), the strain ($\eta$), 
and the stress ($\sigma$).
Usually we apply electric fields $E$ and elastic stresses $\sigma$ externally,
so we can regard the polarization and the strain as "internal" or dependent 
variables.  A fundamental postulate of thermodynamics applied to a 
ferroelectric is that its free energy
${\cal F}$ can be generally expressed as a function of ten variables
(three components of polarization, six components of the stress
tensor, and finally one of temperature). Our goal here is to write
down an {\em ansatz} for this free energy, using symmetry arguments whenever
possible to pare down the number of terms. The second important
thermodynamic principle that we will also employ is that the values 
of the dependent variables in thermal equilibrium are obtained at the 
free energy minimum when the free energy is optimized.

We make the key approximation that in the vicinity of
a phase transition we can expand the free energy in powers of the dependent 
variables with coefficients that can be fit to experiment or gleaned
from microscopic calculations.
In the best case, we may be able to truncate this series with only a few terms.
In order to be more specific, let us take a simple
example where we expand the free energy in terms of a single component
of the polarization; for the moment we ignore the strain field, an assumption
which might be appropriate for a uniaxial ferroelectric. We shall choose the
origin of energy for the free unpolarized, unstrained crystal to be zero,
and hence we write
\beq
\label{freeP}
{\cal F}_P = \frac{1}{2} a P^2 + \frac{1}{4} b P^4 + \frac{1}{6} c P^6 - EP \;\;
\enq
where we have truncated the power series at the sixth term, and
$a$, $b$ and $c$ are coefficients that will be discussed
in more detail shortly.
Here we will use the notational convention that $\cal{F}$ and
$F = \int dV {\cal F}$ refer to the free energy density and the 
total free energy
respectively.
The equilibrium configuration is determined 
by finding the minima of ${\cal F}_P$, where we shall have
\beq
\frac{\partial {\cal F}_P}{\partial P} = 0 \;\;\;.
\enq
This equation gives us an expression for the
electric field $E$ as a function of the polarization
\beq
E = aP + bP^3 + cP^4
\label{E}
\enq
Thus we can determine the linear dielectric susceptibility 
above
the transition by
differentiating this equation with respect to $P$ and then
setting $P = 0$ to obtain
\beq
\chi = \frac{P}{E} = \frac{1}{a}.
\enq
In the Landau-Devonshire theory it is assumed that around
the Curie point ($T \sim T_0$)
\beq
a = a_0 (T - T_0)
\label{a}
\enq
and the other coefficients in the free energy expansion
are independent of temperature.
Combining the last two equations, we find an expression for
the dielectric stiffness
\beq
\kappa = \frac{1}{\chi} = a_0 (T - T_0)
\label{dstiffness}
\enq
which captures the Curie-Weiss behavior (in $\chi$) observed 
in most ferroelectrics for $T > T_0$; this provides
additional support for the linear temperature ansatz for $a$.
For the sake of completeness, we note
that the temperature $T_0$ where $a$ changes sign is close
but not exactly coincident with the Curie temperature, as we
shall discuss shortly.  

%************************************************************************
\begin{figure}
\begin{center}
\includegraphics[height=1.5in]{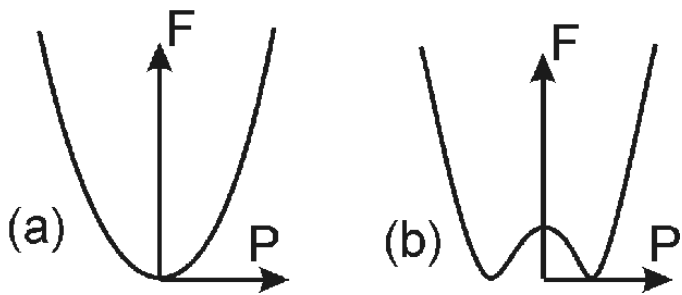}
\end{center}
\caption{\label{fig:freeP} Free energy as a function of polarization for
(a) a paraelectric material, and for (b) a ferroelectric material}
\end{figure}
%***************************************************************************

If we include the linear temperature-dependence of $a$, we have
the general expression for the free energy
\beq
\label{freeP5}
{\cal F}_P = \frac{1}{2} a_0 (T - T_0)  P^2 + \frac{1}{4} b P^4 + 
\frac{1}{6} c P^6 - EP \;\;
\enq  
where, to our knowledge,
$a_0$ and $c$ are both positive in all known 
ferroelectrics \cite{Fatuzzo67,Lines77}.  Figure 2 shows the free 
energy as a function of polarization
in the paraelectric ($T >> T_0$) and the ferroelectric ($T << T_0$)
phases.  How this free energy transforms between these
two configurations will be determined by the sign of the coefficient
$b$; as we shall see shortly, its sign will determine the nature
of the parelectric-ferroelectric transition,
and whether the polarization at $T < T_0$ develops continuously
or discontinuously.

\subsection{Second-Order (Continuous) Transition}

If $b > 0$, then a second-order transition occurs at $T = T_0$,
and the free energy will evolve continuously as a function of
decreasing temperature from the first schematic ($P = 0$) in Fig 2 to the
second, that has minima at finite polarizations $P = \pm P_0$. 
The spontaneous polarization can be estimated by setting $E=0$ in
(\ref{E}); since all the coefficients are positive, we will only
retain the two lowest-order terms.  The result is
\beq
P_0 = \left\{\frac{a_0}{b} (T_0 - T) \right\}^{\frac{1}{2}}
\label{Ps}
\enq
where we see that the spontataneous polarization $P_s$ will increase
with decreasing temperature from the point $T = T_0$.
We note that if we determine the dielectric stiffness below
the transition ($T < T_0$) then we find
\beq
\kappa = 2 a_0 (T_0 - T)
\label{dstiffness2}
\enq
which is to be compared with ({\ref{dstiffness}), its value just
above $T_0$; perusal of these two expressions suggests that $\kappa$ vanishes
at $T = T_0$ and that consequently the dielectric suseptibility
diverges.  In practice, 
the latter
achieves large but finite values in real materials where this expression
is appropriate \cite{Fatuzzo67,Lines77}.
We can now also solve for the discontinuity in the specific heat at
the transition; using $P = 0$ for $T > T_0$ and (\ref{Ps}) for $T < T_0$,
we determine 
\beq
\Delta C_v = C_v (T = T_0^+) - C_v (T = T_0^-)
\enq
where $C_v \equiv - T \frac{\partial^2 {\cal F_P}}{\partial T^2}$
to obtain
\beq
\Delta C_v = \frac{a_0^2 T_0}{2 b}
\label{cv}
\enq
for the Landau expression for this quantity.
Schematics associated with this second-order phase transition
as described by Landau-Devonshire theory are displayed in
\figref{fig:2ndorder}. 

%**********************************************************************
\begin{figure}
\centerline{\includegraphics[height=2.5in]{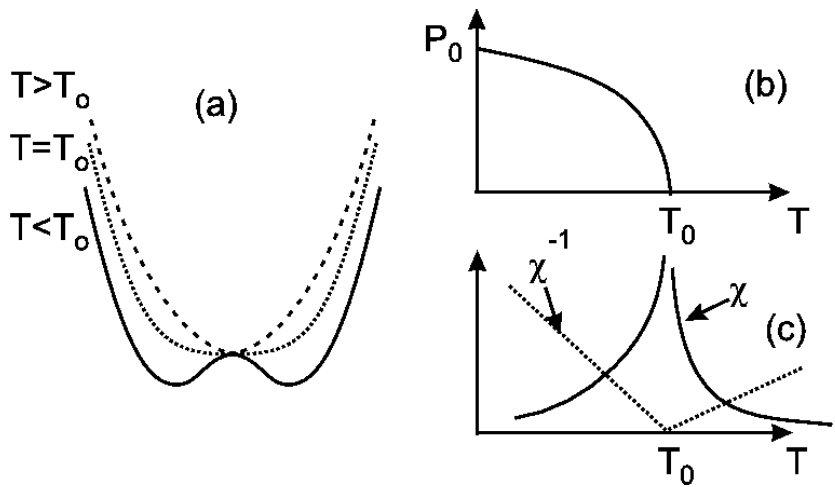}}
\caption{\label{fig:2ndorder}Second order phase transition.
(a) Free energy as a function of the polarization at $T>T_0$,
$T=T_0$, and $T<T_0$;
(b) Spontaneous polarization $P_0(T)$ as a function of temperature
(c) The susceptibility $\chi$ and its inverse, where
$\chi = \partial P/\partial E |_{P_0}$ is evaluated at the equilibrium
polarization $P_0(T)$ }
\end{figure}
%**********************************************************************

\subsection{First-Order (Discontinuous) Transition}

Logically (and practically as it turns out), we should also consider the case
of $b<0$ (while c remains positive). This is sketched in
\figref{fig:1storder}.

%***********************************************************************
\begin{figure}
\centerline{\includegraphics[height=2.5in]{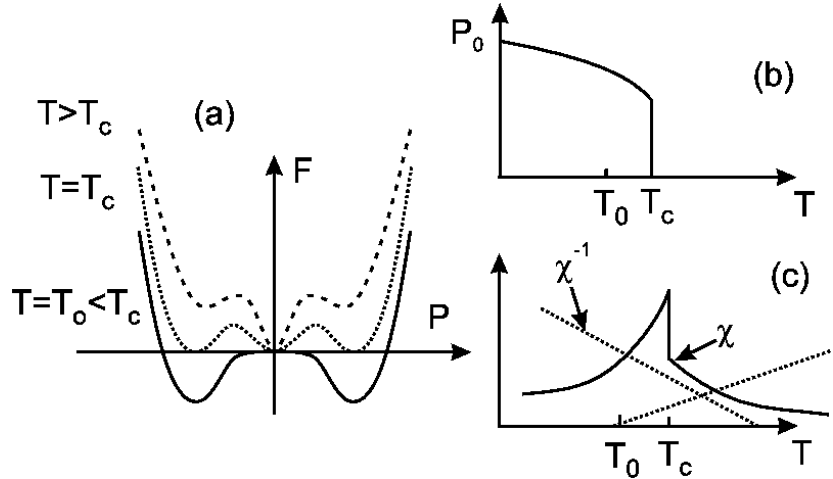}}
\caption{\label{fig:1storder} First order phase transition.
a) Free energy as a function of the polarization at $T>T_c$,
$T=T_c$, and $T=T_0<T_c$;
(b) Spontaneous polarization $P_o(T)$ as a function of temperature
(c) Susceptibility $\chi$. }
\end{figure}
%**************************************************************************

With the negative quartic coefficient, it should be clear that even
if $T>T_0$  (such that the quadratic coefficent is positive) the free energy
may have a subsidiary minimum at non-zero $P$. As $a$ is reduced, which
corresponds physically to the reduction of the temperature,
this minimum will drop in energy 
below that of the unpolarized state, and so will be the thermodynamically
favored configuration. The temperature at which this happens is, by 
definition, the Curie temperature $T_c$ which however now
exceeds $T_0$. At any temperature between $T_c$ and $T_0$ the
unpolarized phase exists as a local minimum of the free energy.
The most important feature of this phase transition is that the
order parameter jumps discontinuously to zero at $T_c$. This type
of phase transition is usually called a first-order or discontinuous
transition, of which solid-liquid transitions are common examples.

The procedure for finding the spontaneous polarization and the
linear dielectric susceptibility is conceptually the same as
before, but now one cannot neglect the quartic and sixth order
terms. Qualitatively we then find a dielectric stiffness (inverse
of the linear susceptibility) that does not vanish at $T_0$,
corresponding to the finite jump in both the susceptibility
and the spontaneous polarization at the transition. Schematics
of the free energy, the spontaneous polarization, the dielectric
stiffness and the linear susceptibility are shown in \figref{fig:1storder}.
We note that at $T=T_c$ the three minima are energetically degenerate.
As a result, the system's behavior at $T=T_c$ will depend on whether it
is approaching $T_c$ from lower or higher temperatures.  More specifically,
the system will be in one of the two finite polarization ($P \neq 0$)
minima if it is heated from an initial low temperature $T_i < T_c$, whereas
it will be in a paraelectric state ($P = 0$) if the initial temperature
is high ($T_i > T_c$).  Indeed
the phenomenon of thermal hysteresis, where the transition temperature
depends on whether the sample is heated or cooled, is prevalent
in a number of first-order ferroelectrics including barium titanate \cite{Fatuzzo67}.
We emphasize that 
it is only for $T_0 < T_c$ that the ferroelectric minima are 
thermodynamically favorable.

\begin{figure}
\begin{center}
\includegraphics[height=3in]{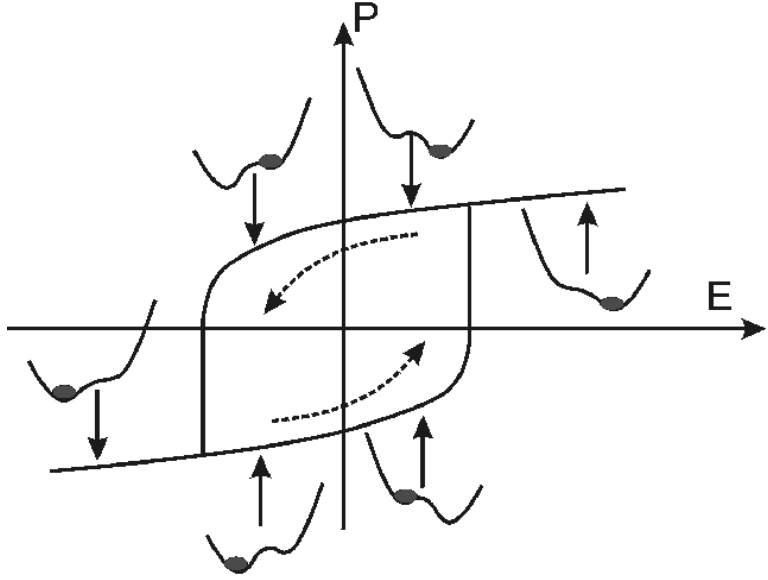}
\end{center}
\caption{\label{hysteresis} Schematic picture of hysteresis in an idealized
ferroelectric}
\end{figure}

In a ferroelectric below $T_0$ there are (at least) two minima of the 
free energy, corresponding to spontaneous polarizations of different 
spatial orientations. 
The barrier between these minima means that a small electric field will 
not immediately switch the polarization. We note that the Landau-Devonshire 
theory described here
predicts hysteresis \cite{Fatuzzo67,Lines77}, 
shown schematically in Figure 5, 
in the ideal (and fictitious) case where all the 
dipoles have to be overturned together to switch
from one polarization orientation to the other.

We note that for ferroelectrics with either continuous or discontinuous
transitions, values of the Landau coefficients can be determined by comparing
experimental measurements of $\kappa$ and $P_s$ to the Landau-Devonshire
expressions; such parameter sets for a number of commonly studied
materials are included in an Appendix at the end of this chapter.
 
\subsection{Coupling to strain}

An important feature of ferroelectric materials is their great
sensitivity to elastic stress. In order to understand why this is so, we
can again take recourse to Landau theory by adding in strain
dependent terms to $(\ref{freeP5})$.
The strain in a solid is measured by how the displacement
$\vec u$ of a point in the solid varies with position $\vec r$, and
since this is the dependence of a vector upon a vector, the answer is
a tensor: the strain is usually defined as
\beq
\label{eq:straindef}
\eta_{ij} = \frac{1}{2} \left(\frac{\partial u_i}{\partial r_j}
+  \frac{\partial u_j}{\partial r_i}  \right) \;\;\;
\enq
where here $i,j$ indicate the $x,y,z$ components of the vectors.
$\eta$ is therefore a 3x3 symmetric matrix, with six independent
components. In materials that are cubic (or nearly so) there
will be three independent components to the strain:  the
volume strain (uniform in all three directions and two kinds
of shear). 

In general, the polarization will couple to one or more types of
strain, and specifically which ones
can generally be seen by inspection. Consider
a cubic crystal (e.g. $BaTiO_3$) that undergoes a ferroelectric
phase transition to a state where the polarization can point along one
of the six orthogonal cubic directions. Now it is clear that there
is a special axis (one of the six directions after the symmetry
has been broken) and so it would no longer be expected that the
crystal as a whole will remain cubic --- one expects a distortion
into a tetragonal crystal, which can be described by a tetragonal
strain $\eta$.
The fact that the lowest order coupling allowed in this case is of
the form $\eta P^2$ (and not, for example, $\eta P$ or $\eta^2P$) 
is of course a matter of
symmetry. 

For a uniaxial ferroelectric, the leading order terms will be of the
following form
\beq
\label{fstrain}
{\cal F}_\eta = \frac{1}{2} K \eta^2 + Q \eta P^2 + ..  - \eta \sigma
\;\;\;,
\enq
Here $\eta$ is (a component of) the strain field, and the first term
represents Hooke's law, namely that the elastic energy stored in a
solid is quadratically dependent on the distortion, so $K$ is
(one of) the elastic constant(s). The second term is a coupling
between the elastic strain and the polarization; the fact that
this is linear in the strain and quadratic in the polarization
depends on the special symmetry of the transition. 
While this is the leading term in pseudocubic materials,
there are other materials (e.g. $KH_2PO_4$) where the symmetry is lower, 
and the coupling can be of the form $\eta P$ -- linear in both strain and 
polarization. Materials with a linear relation between stress and 
polarization are called piezoelectric.

Using the free energy, which now consists of the terms in
$(\ref{freeP5})$ and $(\ref{fstrain})$,
${\cal F} = {\cal F}_P + {\cal F}_s$. we can now determine the
properties in equilibrium by minimizing with respect to both $P$ and
$\eta$, viz
\beq
\frac{\partial {\cal F}(P,\eta)}{\partial P} =
\frac{\partial {\cal F}(P,\eta)}{\partial \eta} = 0 \;\;\;
\enq
Let us take the second of these equations first:
\beq
\frac{\partial {\cal F}(P,\eta)}{\partial \eta} = K\eta + Q P^2 - \sigma
\enq

There are a few different limits to look at.
Firstly, note that if the polarization is zero, we get Hooke's law
$\eta = \sigma / K$.
The second -- apparently trivial case -- is when a stress is applied
to force the strain to be exactly zero at all times. This is not as
absurd as it seems, because often crystals can be considered to
be clamped by their surroundings so that no strain is allowed at all.
One common situation is of a thin epitaxial film which is forced to
have the lattice constants matched to the substrate, and is free to relax
only in the third direction.
In the case of perfect ``clamping'' $\eta =0$, and the free energy is just
as before.

The third case to consider is when no external stresses are applied
($\sigma = 0$), and we then have
\beq
\eta = -\frac{Q P^2}{K}
\enq
so that a spontaneous (tetragonal strain) occurs proportional to the
square of the polarization. Notice now that we can substitute for the
strain as a function of polarization, and we have a free energy
\beq
{\cal F}(P,\eta(P)) =
\frac{1}{2} a P^2 + \frac{1}{4} (b-2Q^2/K) P^4 + \frac{1}{6} c P^6
+ ... - EP \;\;.
\enq

In comparison with the clamped system, the only change is to
reduce the quartic coefficient (notice that the result is independent
of the sign of $Q$). This means that in the case of an already
first-order transition ($b<0$) the transition is driven even more
strongly first order, and $T_0$ is raised.  In many ferroelectrics, 
the effects of clamping can be large - shifting the
transition temperature by tens of Kelvin, and even changing the order 
of the transition. If
$2Q^2/K >b>0$,  a first-order transition becomes
second-order in a clamped system where the strain is allowed to relax;
this is the case \cite{Fatuzzo67} in $BaTiO_3$.

Application of external hydrostatic pressure leads to a 
shift \cite{Merz50} in $T_0$ whose sign is determined by 
that of the volume expansion coefficient;
a reduced lattice constant is accompanied by a reduction in $T_0$, consistent
with measurements where similar effects have been achieved by chemical
substitution \cite{Merz50}. The effect of biaxial stress, where the 
crystal contracts
along two axes and expands along the third, on $T_0$ was also studied
both theoretically and experimentally \cite{Forsbergh54}; enhancement 
of $T_0$ was predicted 
and observed in barium titanate crystals \cite{Forsbergh54}.  We shall see 
shortly that this effect is even more dramatic in thin films where such 
elastic effects occur at the interface to the substrate.

\subsection{Domains}
     So far we have pretended that the polarization in a ferroelectric
     can be treated as entirely uniform, and this is far from the case.
     There are many reasons for the existence of domains, including non-uniform
     strain, microscopic defects, and the thermal and electrical history 
     of the
     sample. But even in an ideal crystal, domains are to be
     expected for energetic reasons associated with electrostatics.

     The macroscopic bulk polarization is produced by a displacement of
     positive charge relative to negative charge; at the surface of the sample
     there must then be a net charge density of opposite signs on the opposite
     sides of the crystal (\figref{fig:surfpol}).
     \begin{figure}
     \centerline{\includegraphics[height=1.4in]{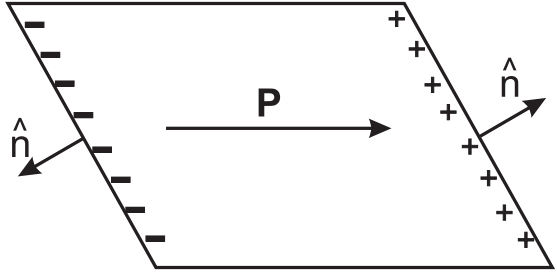}}
     \caption{\label{fig:surfpol} Surface charge density generated by
     a bulk polarization at an interface}
     \end{figure}
%% New insertion 19/8/06
Any inhomogeneity in the polarization acts like a charge density - and in particular the discontinuity in the polarization at an interface or boundary acts like a surface charge, as can be seen by the following elementary electrostatic argument.

The potential $V( \vec r )$ induced by a dipole $\vec p$ at the origin is
\beq
V( \vec r ) = - \frac{1}{ 4 \pi \epsilon_o} \vec p \cdot \vec \nabla \left ( \frac {1}{r} \right ) \;\;,
\enq
and thus the potential due to a distributed polarization $\vec P ( \vec r )$ in a volume 
$\tau$ bounded by a surface $\vec S$ is
\beq
V( \vec r ) =   \frac{1}{ 4 \pi \epsilon_o} \int _{\tau} d {\vec {r^\prime}} \;
{\vec P } ({\vec {r^\prime}}) \cdot {\vec {\nabla_{r^\prime}}} \left ( \frac {1}{|{\vec r} - 
{\vec {r^\prime}} |} \right ) \;\;.
\enq

Using the product rule: $\vec \nabla \cdot (\vec A f) = f \vec \nabla \cdot \vec A + \vec A \cdot \vec \nabla f $, we have

\begin{eqnarray}
V( \vec r ) &=&  \frac{1}{ 4 \pi \epsilon_o} \int _\tau d {\vec {r^\prime}}
\left [ 
- \frac {\vec \nabla \cdot \vec P ({\vec {r^\prime}}) } {|{\vec r} - {\vec {r^\prime}} |}
+ \vec \nabla \cdot \left ( \frac {\vec P  ({\vec {r^\prime}}) }
{ {|{\vec r} - {\vec {r^\prime}} |}  } \right )
 \right ], \\
&=& \frac{1}{ 4 \pi \epsilon_o}  \left [ -\int _\tau d {\vec {r^\prime}}
\frac {\vec \nabla \cdot \vec P ({\vec {r^\prime}}) } {|{\vec r} - {\vec {r^\prime}} |} 
+ \int_{\vec S} \frac {d \vec S \cdot \vec P ({\vec {r^\prime}}) }
{|{\vec r} - {\vec {r^\prime}} |} \right ] \;,
\end{eqnarray}
where the last manipulation uses the divergence theorem leading to a surface integral.

The potential appears to be generated by a distributed bulk charge density 
\beq
\rho (\vec r ) = - \vec \nabla \cdot \vec P (\vec r )
\enq
as well as a  surface charge density 
     \beq
     \sigma = \vec P \cdot \hat n
     \enq
     where $\hat n$ is the vector normal to the boundary surface $\vec S$. Occasionally, some books that like to make arbitrary distinctions between ``bound'' and ``free'' charge will call the charges generated by an inhomogeneous polarization ``fictitious''. These charges are fictitious only in the sense that they are conserved - namely if we begin with an electrically neutral system and create a polarization density from it by moving the electrons relative to the ions, the integral of the polarization charge (over the whole system) must vanish. They are in
all other respects quite real. The surface charges
     generate electric fields, both internal and external
     to the sample, and the fields themselves store energy - just as in a
     capacitor. In an approximation where the polarization $\vec P$ is assumed to be homogeneous within a domain, and changing abruptly at surfaces or interfaces, only the surface term survives.

     The system will minimize its energy by eliminating -- as far as is possible
     -- the surface charges, and in a thin film for example this makes clear
     that the preferred orientation of the polarization will be
     in the plane of the film, rather than pointing perpendicular
    to the film. If one has a crystal that is thick in all dimensions, another
    situation is preferable -- to introduce domains where the polarization
    is oriented to be always parallel to the crystal surface.
    Such a situation can of course only be achieved by introducing
    domain walls into the bulk, and this will produce polarization charges
    unless the walls are appropriately oriented.

     The interface charge density between two neighbouring domains is, by
     extension of the formula at a free surface,
     \beq
     \sigma = (\vec P_1 - \vec P_2) \cdot \hat n
     \enq
     where $\vec P_1$, $\vec P_2$ are the polarizations of the two domains,
     and $\hat n$ is a unit vector normal to the interface.
     The two cases when the surface charge will exactly vanish are
     when the polarizations are antiparallel to each other, and parallel
     to the domain wall (called a $180^\circ$ domain wall), or
     when the domain wall bisects the angle between two domains
     pointing head-to-tail. Because we are often dealing with nearly
     cubic crystals where the possible polarizations are at $90^\circ$
     to each other, the latter is often termed a $90^\circ$ domain
     wall. Both of these wall types  are illustrated in the sketch
     \figref{fig:domain}, which is the ideal configuration of the
     polarization in a `` bar'' of ferroelectric single crystal ---
     and which bears obvious comparison to the ideal magnetic configuration
     in a single crystal bar magnet, that may be familiar to some.

     \begin{figure}
     \centerline{\includegraphics[height=2in]{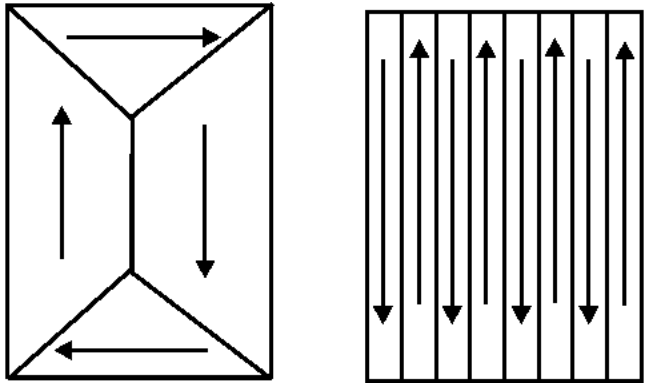}}
     \caption{\label{fig:domain} Ideal domain configuration in a
     single crystal of cubic ferroelectric material, where the coupling
     to strain is negligible. On the right is the configuration
     adopted when strain effects are important. }
     \end{figure}

     The presence of domain walls involves other energy costs. Firstly,
     the domain wall is microscopically different from the bulk, and the
     energy gain of forming the polarized ground state has been lost.
     Secondly, the polarization is coupled to elastic strain -- as we saw above
     -- and we must also make sure that the strain fields are compatible.
     There is no difficulty with a $180^\circ$ domain wall, but a $90^\circ$
     domain wall produces problems, and the ``ideal'' configuration
     of \figref{fig:domain} is by no means ideal in terms of the strain
     fields. The existence of domains in a sample automatically generates
     inhomogeneities in the strain which interact in complex ways.

%%% Elastic compatibility discussion introduced  19/8/6

Some contemplation of the two-dimensional pictures above makes it evident that only certain kinds of domain structures are allowed, unless dislocations, cracks, or voids are introduced into the crystal. The moment we have inhomogeneous strain, we must remember to maintain the condition that the displacement field $\vec u ( \vec r )$ must be {\em continuous}; however, the free energy depends only on the strains $\eta$, which are derivatives of $\vec u$ (see \eqref{eq:straindef}). Imposing the continuity constraint gives rise to the additional (St Venant) compatibility conditions on the strain fields \cite{Love44}
\beq
\label{eq:stvenant}
\vec \nabla \times (\vec \nabla \times \eta ) = 0 ,
\enq
which formally reduces the number of independent strain components. When dealing with the allowed arrangements of domains of (locally) homogeneous strain (except at the boundary), the application of condition \eqref{eq:stvenant} (typically by taking a line-integral around the intersection of three domains) determines the legality of various domain arrangements in two and three dimensions. 

This condition can also be used to integrate out a single component of the strain, at the expense of introducing long-range forces\cite{Kartha95}, and this approach can then be used to support calculations and simulations of domain structures in two \cite{Shenoy99} and three \cite{Rasmussen01} dimensions. This methodology is now quite far advanced in understanding  ferroelastic effects on domain structures in martensites \cite{Lookman03} though it has not yet been combined with simulations of both the elastic and the ferroelectric order parameter.

     There is another feature of ferroelectric domain walls which is not shared
     by magnetic domains. As far as we are aware, magnetic monopoles do
     not exist, and therefore the (fictitious in the sense of conserved) magnetic charges that
     are generated on magnetic boundaries cannot be screened. In ferroelectrics,
     the surface polarization charges can be, and are, screened by
     real electrical charges -- from impurities, defects, and migrating
     ions for example. The motion of charged species (on a
     fairly slow time scale) to domain walls then provides a mechanism
     for memory and associated physical phenomena.

     This brief introduction to domain structure highlights several issues. 
	The first is that we need to extend the Landau-Devonshire theory of a uniform
	ferroelectric to incorporate non-uniform polarization. This we do by a 
parameterization of the free energy now including terms in powers of gradients - this is Ginzburg-Landau theory, introduced in the next section. Secondly, we noted that the 
geometry of the sample influences the domain structure, and in particular that there is
potentially antagonism between the strain fields and electric fields introduced by the domain structure. 
The final major point is to note that the effect of inhomogeneities in ferroelectrics is to introduce {\em long-range} forces, from either charge inhomogeneity or strain fields. It is these two phenomena that complicate a local description of the ferroelectric free energy.

%\newpage

\section{Landau-Ginzburg Theory}

\subsection{General Considerations}

The Landau-Devonshire theory that we discussed in the previous Section
is well-suited for the description of a {\sl poled} bulk ferroelectric 
near its  transition ($T_0$) with a polarization 
for $T < T_0$ that is spatially uniform. However we also offered
physical motivation for treating spatially nonuniform 
polarization in unpoled bulk ferroelectrics, and this will be 
the subject of our treatment in this Section. Generally speaking,
Landau-Ginzburg (LG) theory incorporates small spatial
variations of the order parameter (here the polarization) within
the phenomenological Landau-Devonshire theory with 
assumptions used originally by Ornstein-Zernike; there fluctuations
of the order parameter at different wavevectors are assumed uncorrelated
and thus independent.

Let us now discuss specifics of this Landau-Ginzburg approach.
Slow variations in the direction of the polarization lead to an
additional  contribution
in the free energy density ($\Delta {\cal F}_{LG}$)
that is proportional to $|\nabla P|^2$.
Though here, as before, we restrict our discussion to the case of
a single-component order parameter, we note that this specific quadratic
form of $\Delta \cal F_{LG}$ is valid for more
general vector order parameters based on symmetry considerations
\cite{Landau59,Toledano87}.
Let us return to the scalar order parameter.  Then
if we define a polarization 
density $P(r)$ where $\vec{r}$ is a 
$d$-dimensional spatial vector, then to lowest order
the Landau-Ginzburg free energy is
\beq
F = a_0 (T-T_0) \int d^d \vec{r} \left\{P(\vec{r})\right\}^2 + 
\gamma \int d^d \vec{r} \left\{\nabla P(\vec{r})\right\}^2
\label{lgfe}
\enq
where the second term is simply the leading contribution 
in an expansion of polarization-polarization interactions that estimates
the additional free energy cost if the polarizations at 
different spatial positions are not parallel.
Using the fact that the polarization is real (i.e. $P^*(\vec{q}) = P(-\vec{q})$), then we can perform a standard Fourier transform to rewrite the
Landau-Ginzburg free energy as
\beq
F = \int \frac{d^d \vec{k}}{(2\pi)^d} (a_0 (T-T_0) + \gamma k^2) |P(\vec{k})|^2.
\enq
We note that here we have two quadratic degrees of freedom; by the 
equipartition theorem we can then write
\beq
(a_0 (T-T_0) + \gamma k^2) |P(\vec{k})|^2 = k_B T
\label{equipartition}
\enq
where $k_B$ is the Boltzmann constant.

\subsection{The Polarization Correlation Function}

We are now in a position to determine the static two-site
polarization correlation function
\beq
g(\vec{r}) = \langle P(\vec{r})P(\vec{0})\rangle  - 
\langle P(\vec{0})\rangle^2
\enq
which is a measure of the fluctuations of the polarization.  
We note that in ferroelectrics
such quantities can be measured with diffuse X-ray scattering 
\cite{Lines77,Fujii71}. Taking
a Fourier transform of $g(\vec{r})$ above in the paraelectric phase
(where $\langle P(\vec{0})\rangle = 0$) and assuming that the different
wavevector modes are uncorrelated, we obtain 
\beq
g(\vec{q}) = \langle |P(\vec{q})|^2 \rangle
\enq
which then, combined with the expression in $(\ref{equipartition})$,
yields
\beq
g(\vec{q}) = \frac{k_B T}{a_0 (T - T_0) + \gamma k^2}
\label{g1}
\enq
as the expression for the Fourier transform of the correlation function.
Now we simply take the inverse transform to find
\beq
g(\vec{r}) \sim 
\frac{k_B T}{\gamma} \frac {e^{-r/\xi}}
{r^{d-2}} \qquad T \neq T_0
\label{gr}
\enq
and
\beq
g(\vec{r}) \sim \frac{k_B T}{\gamma} \frac{1}{r^{d-2}}    \qquad T = T_0 
\label{grd}
\enq
where the correlation length is
\beq
\xi = \sqrt{\frac{\gamma}{a_0 |T - T_0|}} 
= \sqrt{\frac{\gamma}{a_0 T_0 }} 
\sqrt{\frac{T_0}{|T - T_0|}}   \equiv \xi_0 |t|^{-1/2}.
\label{xi}
\enq
and $t = \frac{ T - T_0}{T_0}$ is the reduced temperature.
Physically for $T > T_0$ this correlation length, $\xi$, corresponds to the 
length-scale
over which the polarization exists; it diverges at $ T = T_0$, and for 
$T < T_0$ it is associated with the length-scale over which $P$ varies 
from its equilibrium value.  
We note that the polarization correlation function is exponentially decreasing
above the transition, but has a dimensionally-dependent power-law form
when $T \sim T_0$.  As a final note to this subsection, we remark that the 
dimensions of $\gamma$ and $a_0 (T - T_0)$ are $\frac{E}{P^2 L}$
and $\frac{E}{P^2 L^3}$ respectively, so that 
$[\xi_0] = \sqrt{L^2} = L$ as expected.

\subsection{The Levanyuk-Ginzburg Criterion}

How reliable is this Landau-Ginzburg theory, and when is it no longer
valid?  The LG approach is a long-wavelength description of a system 
near a phase transition, where its modes are coarse-grained on scales 
of order the correlation length $\xi$.   The resulting effective
free energy density is written as an expansion of the order 
parameter (e.g. $P$) averaged over a volume $\Omega_\xi$ determined by $\xi$. 
When are the fluctuations of the order parameter,
averaged over the correlation volume $\Omega_\xi$, small 
in comparison with its coarse-grained average?
The answer, relevant for the validity of this long-wavelength approach,
is dependent on system dimensionality.
We have just seen that the 
LG approach 
yields a two-site correlation function, a measure of the order parameter 
fluctuations, that increases dramatically at $T \sim T_0$; it is therefore 
clear that in the immediate vicinity of $T_0$ the LG expansion 
is no longer valid.
So how close can we get
to the transition itself before this approach no longer works?

As it turns out, we can use Landau-Ginzburg theory itself to determine
its own fallibility \cite{Levanyuk59,Ginzburg60}.  Before we present this
argument, let us return to Landau-Devonshire theory and look at its 
predictions for behavior near the phase transition.  
We recall the expression for the free energy
\beq
\label{freeP2}
{\cal F}_P = \frac{1}{2} a_0 (T - T_0)  P^2 + \frac{1}{4} b P^4                     - EP \;\;
\enq  
which we have truncated at quartic order.  The equilibrium value
of the polarization corresponds to a free energy minimum
\beq
\frac{\partial {\cal F}_P}{\partial P} = a_0 (T - T_0) P + b P^3 = 0
\enq
which yields (for $ T< T_0$)
\beq
P = \sqrt{\frac{a_0 (T_0 - T)}{b}} \sim (-t)^{1/2}.
\label{PP}
\enq
whereas $P=0$ in the paraelectric phase.  The form of the free energy, (\ref{freeP2}), combined with these results for the polarization imply that the 
associated specific heat, 
$C_v = - T \frac{\partial^2 F}{\partial T^2}$, has a discontinuity
at the transition
so that the mean-field exponent $\alpha$ is zero (i.e. $C_v \sim |t|^0$).

Equipped with this information, we are now ready to determine
the condition for when Landau-Ginzburg theory breaks down.  Basically
this occurs when the fluctuation free energy associated with a typical
fluctuation of order the correlation length is comparable to the total
free energy.  The fluctuation free energy per unit volume can be estimated 
as
\beq
{\cal F}_{fluct} \sim \frac{kT}{\xi^d} \sim |t|^{\nu d}
\enq
where we have used $\xi \sim |t|^{-\nu}$.  We have defined the specific heat,
$C_v \sim |t|^{-\alpha}$ so that two integrations yield ${\cal F} \sim |t|^{2- \alpha}$.
Thus for a consistent theory, for $t \rightarrow 0$, we must have
\beq
d^* > \frac{2 - \alpha}{\nu}.
\enq
Inputting the mean-field values ($\alpha = 0$ and $\nu = 1/2$), we get
\beq
d^* > 4
\enq
which indicates that above the upper critical dimension, $d^* = 4$, the LG  
approach is valid.  We note that the value of $d^*$ depends
on the form of the Landau expansion and can be different for systems
with different underlying symmetries \cite{Als-Nielsen77}.

This calculation indicates that for real materials in $d < 4$,
the Landau-Ginzburg approach breaks down close to the phase transition...but
how close?  Intuitively it seems that the larger the number of neighbors the
better it does, which suggests that it works better for systems with
long-range forces.  Let us use what we have learned to put this speculation
on firmer footing, and here we follow the original Levanyuk-Ginzburg 
reasoning \cite{Levanyuk59,Ginzburg60}.  For $T < T_0$ let us argue that 
coarse-grained fluctuations
in the polarization must be small in comparison
with the average polarization itself; this translates into the condition
\beq
g(\vec{r}) \ll  P^2
\label{lgcondition}
\enq
where $|r| \sim \xi$. 
Using our previous results in $(\ref{gr})$ with $d=3$ and $(\ref{PP})$, 
we can rewrite
this expression as
\beq
\frac {k_B T_0}{\gamma \xi(T)}  \ll \frac{a_0 (T - T_0)}{b}
\enq
which, when using specific expressions for $\xi(T)$ and $C_v$ in
$(\ref{xi})$ and $(\ref{cv})$, 
results in the standard form of the Levanyuk-Ginzburg criterion
\beq
{\cal A} \frac{1}{(\Delta C_v)^2 \xi_0^6}  \ll |t|
\label{Ginzburg}
\enq
where ${\cal A}$ is a constant that is unimportant for our present purposes; 
here the key point is
that the range of the interaction, $\xi_0$, plays an important role
in determining the validity of the mean-field theory. More specifically,
the Levanyuk-Ginzburg temperature,    
determined by the expression in $(\ref{Ginzburg})$, is inversely proportional
to the {\sl sixth} power of the range of the interaction, and thus
Landau-Ginzburg theory should be quite reliable in
the vicinity of the paraelectric-ferroelectric transition.  
We note that if we generalize the 
condition (\ref{lgcondition}) to d-dimensions then, using (\ref{grd}) with
$|r| \sim \xi$  and 
(\ref{PP}), we obtain
\beq
{\cal B} |t|^{\frac{d-2}{2}} << |t|^1
\enq
which is satisfied for $d > d^*= 4$ for arbitrarily small $|t|$ 
where ${\cal B}$ is a constant.  We have thus
recovered the same result that we obtained earlier in this section, namely
that 
Landau-Ginzburg theory is exact for dimensions greater $d > 4$ for the 
ferroelectrics under consideration.  Here we remark that the 
angular-dependence of the dipolar interactions has not been considered 
here, and it is indeed this feature that leads to logarithmic corrections
to mean-field exponents in certain
experimentally observable (i.e. $d = 3$) cases \cite{Larkin68}.
We note that here we have
referred to this condition as the Levanyuk-Ginzburg criterion
(rather than simply that due to Ginzburg) since we have learned \cite{Dima}
that Levanyuk, a student of Ginzburg's, derived this condition
independently of his advisor and indeed published the result
in a sole author publication \cite{Levanyuk59} a year before 
Ginzburg's paper appeared \cite{Ginzburg60}.
  
\subsection{Displacive and  order-disorder  transitions}
% section added by PBL 3/17/06, modified 10/08/06
The Levanyuk-Ginzburg criterion indicates the validity of the mean-field 
approach to the transition,
but more fundamental in many practical senses is the validity of the continuum approximation
underlying the LG theory. A continuum theory is of course good close enough to a second
order critical point since the diverging correlation length $\xi $ is much larger than the interatomic spacing
--- this is the condition that coarse graining works. But {\em far} from the transition --- either well above $T_c$
or well below it --- this condition will not necessarily hold.

If one takes the Landau theory literally as a description of the phase transition and its dynamics, the order parameter
vanishes uniformly everywhere above the transition temperature. Equivalently one would say that the lattice displacements
corresponding to the electronic polarization are equally uniform - and zero above $T_c$, finite below. Such a type of transition is termed {\em displacive}. A contrasting limit for an ordering transition is that of local moment magnetism \cite{White79}, wherein the magnetic moment on an atom persists above $T_{curie}$ but the long-range order is destroyed. 
In the case of a ferroelectric (or other structural transition), this limit is described as an {\em order-disorder} transition, where the local symmetry-breaking distortions are present in every unit cell above $T_c$, but are randomly oriented at high temperatures, so there is no net polarization. Any real material will of course lie somewhere along the 
spectrum between these two limits.

We have argued above that one reason for the practical usefulness of the LG approach is that there are intrinsic interactions (elastic and charge)
that are long-range, and these induce a long correlation length. But there are also conditions related to the underlying microscopic drivers for ferroelectricity that are ineluctably electronic in nature.
Although in this chapter we are largely avoiding microscopic theory here we engage in
a short digression on the topic of {\em electronic ferroelectricity}, which has some relevance to the issue. 

Spontaneous development of a ferroelectric moment is always associated with a (broken symmetry) atomic displacement,
and often this is pictured as an essentially rigid displacement of charge. But of course the driving force for ferroelectricity
arises from interactions in the electron system, so a completely rigid displacement of ions is not to be expected. One may study this even in linear response theory - how much charge redistribution is produced by a small displacement of the ion - or equivalently how large an electrical dipole is produced by a particular phonon displacement. Phonons in solids can be separated into acoustic and optical branches, where the eigenmodes of the latter describe the relative motion of different ionic species against each other.
One may quantify the electronic rigidity for an optic mode by a number known as the transverse electric charge $Z_T^*$ \cite{Born54}, which measures (to linear order)
the average electrical dipole moment per unit cell $p$ generated by the corresponding relative ionic displacement $u$ (taken to be the same in every cell), viz.
\begin{equation}
p =  Z_T^*e u
\end{equation}
where e is the electron charge. There is a different value of effective charge $Z_T^*$ for each optical mode ( and it may be zero if the displacement preserves inversion symmetry); even a non-ferro-electric crystal will usually have dipole-active modes. A simple case is just that of the rocksalt structure, where the optical eigenmode is a displacement of the two sublattices relative to each other.
The effective charge can be straightforwardly measured by the optical response of an infra-red active phonon. The connection to ferroelectricity is that if such a phonon goes ``soft'' and acquires a finite frozen amplitude $u_0$, then $Z_T^* e u_0$ is the approximate magnitude (to linear order) of the ferroelectric moment per unit cell. This is the classic soft mode theory of ferroelectric instability \cite{Cochran60,Cowley62,Bruce81}.

In the context of the Landau theory, this relates the polarization to the
atomic motion. In a rigid ion picture, one expects $Z_T^*$ to reflect the charge on the ion, but it is in fact not uncommon
to measure effective charges that are quite large: for example\cite{Burstein70} in the rocksalt structure  IV-VI compounds ($Z_T^*$ values in brackets) PbS (4.8), PbSe (5.8), PbTe (6.5), SnTe(8.1), GeTe(11), where a rigid ion viewpoint would lead one to expect  $Z_T^* \le 2$. Only the last two of these materials (which are the least ionic in the Pauling sense!) become ferroelectric at low temperature. 
What these large numbers represent physically is of course that as the nuclei are displaced the electron distribution sloshes from one side of the ion to another, with the electrons moving much further than the nuclei themselves. This large amplification of motion happens
in electronically driven transitions of the Peierls type \cite{Littlewood79}; it can also happen in cases of electronically-driven spontaneous symmetry breaking where the Coulomb interaction favours the breaking of spontaneous orbital symmetry \cite{Portengen96,Batista04}. 

In these cases, the large induced polarization arises because of a broken orbital symmetry. Whatever the underlying model, in what one can see in the not entirely formal limit of a ferroelectric that has $Z_T^* \to \infty$, the transition is indeed electronic in nature. Moreover it corresponds in that case to a metal-insulator transition as well, where the distortion leads to the opening of a small electronic energy gap $2 \Delta$ in the spectrum. Such 
transitions are in fact charge density wave instabilities in disguise, where the broken symmetry happens also to break inversion symmetry, and the CDW onset is also that of ferroelectricity. The driving force for these transitions is the opening (or increase of) an electronic gap at the fermi energy
(for a review, see \cite{Gruner94}).

Suppose for a moment that we can ignore the lattice dynamical effects on this transition, we can then estimate the transition temperature and correlation length in a weak coupling limit, which is well-known to be mathematically just like the Bardeen-Cooper-Schrieffer theory of superconductivity \cite{Schrieffer64,deGennes66,Tinkham75}:
\beq
T_c^{MF} \propto \Delta \simeq W e^{-1/\lambda}
\enq
\beq
\frac{\xi_0}{a} \simeq \frac{W}{\Delta}
\enq
where $W$ is the electronic bandwidth, $a$ the lattice constant, and $\lambda \ll 1$ the effective dimensionless interaction strength (scaled by the electronic bandwidth), which must be small for this kind of theory to be appropriate. In weak coupling, the gap and transition temperature are small (in comparison to the electronic bandwidth) and the correlation length $\xi_0$ is long. 

Now we consider the effect of the coupling to the lattice, and in particular we must take account of dynamical phonons. Provided $k_B T_c^{MF}$ is much smaller than a characteristic phonon energy $\hbar \omega_D$, there will be no appreciable fluctuations of dynamical modes until one is at a temperature very close to the transition, so the mean-field theory is rather good as an estimate of the transition temperature. And moreover, even above $T_c$ the dynamical modes are weakly excited so there are few thermal fluctuations of the lattice. 
All that remains is just the (small, in this case) mean-field lattice distortion that smoothly vanishes at $T_c$. We see that the condition for the displacive limit is that of {\em ultra-weak} coupling, namely that
\beq
\frac{\hbar \omega_{D}}{W} \gg e^{-1/\lambda} \,\,.
\enq
However, if this condition is violated, thermal fluctuations of dynamical modes will become important, and the extra entropy associated with the lattice distortions will drive a transition to the paraelectric phase at a temperature well below the mean-field one \cite{McMillan77}. Here we now crossover to the ``order-disorder'' regime of the transition, and we remark that (as commonly observed in CDW transitions) this crossover happens when $\lambda$ is still very small, since typically the electronic bandwidth is two to three orders of magnitude larger than a typical phonon energy. 

Thus we see that there are two potentially quite distinct reasons to stabilise the displacive picture. One is the coupling to elastic strain, and the presence of long-range forces, about which we will have more to say later. This is important in all classes of ferroelectrics. A second type  is where the transition is predominantly driven by electronic redistribution of states near the fermi level of a metal or narrow-gap semiconductor, and furthermore the displacive limit obtains only when the coupling is very weak, so that electronic gaps are small in comparison to the Debye frequency. These CDW-like systems are rare as classic ferroelectrics, (except maybe for SnTe, which has a transition near 40K and shows the classic soft-mode behavior\cite{Cowley96}), but they are probably quite common in small-moment ferroelectrics where the ferroelectricity is an ``accidental'' adjunct to a charge-ordering transition. Such is potentially the case in charge-ordered manganites, for example \cite{Efremov03}.

\subsection{Recent Developments in Bulk Ferroelectricity}

Landau-Devonshire theory has been successful at reproducing the
observed phase behavior in simple ferroelectric compounds (e.g.
$BaTiO_3$ and $PbTiO_3$) and these results are described in 
many textbooks \cite{Fatuzzo67, Lines77,Jona93}.   Recently it was 
shown \cite{Vanderbilt01} that the standard sixth-order free energy 
expansion can not account for an observed phase where the
polarization is not symmetry-restricted to an axis.
More specifically, a monoclinic phase was observed in 
PZT for a given
temperature and compositional range \cite{Noheda99}, 
that did not emerge from a standard Landau-Devonshire (sixth-order)
treatment \cite{Haun89}. 
However this phase
did emerge in the observed parameter regime from  
simulations \cite{Bellaiche00}
based on a first-principles effective Hamiltonian 
approach \cite{Zhong95}; this result suggested that
no additional physics (or specifically instabilities) were 
necessary for its description. An eighth-order expansion of
the free energy yielded the observed monoclinic phase \cite{Vanderbilt01}
where it was argued that the higher-order terms might be
generated by disorder-averaging.  This analysis will be
relevant for other ferroelectric phases where the polarization
is constrained to a symmetry plane.

In the Introduction we alluded to a link between first-principles 
calculations and Landau theory, and here we  
discuss it more specifically.  In the last two decades, there has been great
progress in the collective understanding of atomic-scale ferroelectricity
through detailed density-functional theory investigations.  We refer the
interested reader to two recent reviews \cite{Ahn04,Dawber05} and other chapters
in this book for more information about these developments. First principles
 effective Hamiltonians,
based on these first-principles approaches with reduced degrees of freedom,
have been analyzed by statistical methods to explore the finite-temperature
behavior of ferroelectrics.  In particular such an effective Hamiltonian
approach was very successful in reproducing the known phase 
behavior \cite{Zhong94} of $BaTiO_3$.  More recently the compatibility of
this particular Hamiltonian and the Landau-Devonshire theory of 
$BaTiO_3$ was demonstrated \cite{Iniguez01} in a careful Monte Carlo study of the order-parameter configuration space.  Thus first-principles density 
functional calculations, via an effective Hamiltonian, can provide the input 
coefficients for Landau-Ginzburg theory; this bridging of theoretical 
methods is crucial for studying phenomena on many length-scales in 
ferroelectrics.

In this Chapter, we focus primarily on the thermodynamic description of 
ferroelectrics using Landau-Ginzburg theory.  However the resulting free energy
can be incorporated into a treatment of thermally activated dynamical 
processes, specifically the determination of nucleation rates \cite{Langer69}.
We note in passing that the coupling between the polarization and the
long-range elastic degrees of freedom in bulk materials implies that 
nucleation of new ferroelectric regions must be a cooperative effect
in that these events cannot occur completely independently.  Dynamical studies
of field-quenched $BaTiO_3$ crystals provide experimental support for this
statement \cite{McWhan85}.  This observed behavior can be qualitatively 
described by a modified time-dependent Landau analysis with simple feedback to
account for the strain-mediation between growing regions \cite{Littlewood86}.

In concluding our discussion on bulk ferroelectricity, we note that there was
much work done in this field in the former Soviet Union that was not 
communicated to the West.  This situation was recently summarized by
Ginzburg \cite{Ginzburg01}, and we point the interested
reader to these thought-provoking reminiscences.  

\section{Reduced Size and Other Boundary Effects}

\subsection{General Discussion}

Ferroelectric materials are very sensitive to electromechanical
boundary conditions due to the long-range nature of their underlying
electrostatic interactions and to the strong coupling between the polarization
and the strain \cite{Scott00,Ahn04,Dawber05}.  Thus the influence of surfaces 
and other boundary conditions on their collective behavior is very 
pronounced \cite{Li96}.
Furthermore there are numerous ways in which the external environment, 
via applied boundary 
conditions, can alter ferroelectric behavior; examples include
electric fields due to surface charges and homogeneous misfit strains.
Finally there is a significant commercial impetus to use ferroelectrics
for portable high-density data storage; for such increasingly miniaturized
applications, key size-dependent effects must be understood and modeled in
order to optimize design \cite{Scott00,Auciello98}.  Typically the fabricated 
structures will be 
of linear dimension 10 nanometer - 1 micron, length-scales that are 
not accessible to
purely atomistic methods.  However such approaches can be used towards
determining the effects of local variations (e.g. compositions, strains, 
displacements) in conjunction with specific boundary conditions; the 
results could then
be incorporated into a bridging phenomenological theory that spans physics
on longer length-scales to enable direct comparison with experimental 
observation.

In this section we summarize key features of Landau-Ginzburg approaches
to ferroelectrics with specific boundary conditions, focusing on the
well-studied case of planar geometries.  We begin with semi-infinite 
ferroelectrics with a free surface, a case that can be easily generalized
to a free-standing thin film.  Because of reduced coordination
number at a free surface compared to the bulk, we expect the average polarization
to display altered behavior at the boundary which could lead to changes in its 
overall thermodynamic behavior \cite{Kretschmer79,Lubensky75}.  We observe 
how this need 
for a specific boundary condition emerges technically from a Landau-Ginzburg approach
of the bulk and the surface free energies.  We also discuss the necessary assumptions 
underlying the emergence of the so-called extrapolation length from such a treatment and 
suggest a test for this often-used (but rarely justified) approach \cite{Ong01}.
If the polarization is normal to the free surface of the ferroelectric film,
a geometry that is often used in practice, then depolarization effects must be included
and this is the topic of the next subsection.  Next we turn to epitaxial misfit strain,
and how it can be turned to dramatically increase the spontaneous polarization of a ferroelectric 
film beyond that in the bulk.  We end with a discussion
of inhomogenous strain and polarization configurations, a more recent
area of investigation that is of particular relevance to three-dimensional 
ferroelectric nanostructures \cite{Scott06a}.

\subsection{The Polarization at the Boundary}

The presence of the Ginzburg term, $|\vec \nabla P|^2$, in the 
Landau-Ginzburg free energy implies that a
boundary condition is necessary to solve the second-order
differential equation that results from the minimization 
procedure.  Let us consider
a semi-infinite
ferroelectric with a second-order phase transition 
We begin by considering the bulk free energy, including the new gradient
term, of
this system
\beq
{\cal F}_{bulk}^{LG} = \int dV \left[ \gamma |\vec \nabla P|^2 - 
\frac{a}{2} |P|^2 + 
\frac{b}{4} |P|^4 \right] 
\label{lgbulk}
\enq
such that $a > 0$ (implying $T < T_0$).
We note that we can use the divergence theorem to split 
the gradient term in $(\ref{lgbulk})$ into a surface and a volume integral
\beq
\int dV ({\vec\nabla} P \vec{\nabla} P) = 
\int dV \left[ \vec \nabla(P \vec \nabla P) - (P \nabla^2 P) \right]
= \int dS (\hat{n} \cdot \vec\nabla P)P - \int dV (P \nabla^2 P)
\label{gradP}
\enq
where $\hat n$ is the normal unit vector to the surface.
Borrowing from analogous studies
of superconducting \cite{deGennes66} and magnetic \cite{Kaganov72,Binder72} 
films, we argue that there is a ``surface tension'' contribution
so that the full free energy
(${\cal F} = {\cal F}_{bulk}^{LG} + {\cal F}_{stension}$)
is now a sum of interior and surface
contributions, ${\cal F} = {\cal F}_{interior} + {\cal F}_{surface}$ ,
where
\beq
{\cal F}_{interior} = \int dV \left[  - \gamma P \nabla^2 P  - 
\frac{a}{2} |P|^2 + 
\frac{b}{4} |P|^4 \right]  
\label{Fint}
\enq
and
\beq
{\cal F}_{surface} = \int dS \left[ \gamma (\hat n \cdot \vec\nabla P) P +
\frac{\alpha (T - T^*)}{2} |P|^2 + 
\frac{\beta}{4} |P|^4 \right]. 
\label{Fsurface}
\enq
We note that this form of the ``surface tension'' free energy contribution
assumes that there exists a temperature $T^*$, not necessarily equal to the
bulk transition temperature $T_0$, where the surface becomes ferroelectric;
here the coefficients $\alpha$ and $\beta$ are usually considered to be 
thickness-independent with $\beta << \alpha$ \cite{deGennes66,Kretschmer79}.

If we vary the free energy density with respect to $\delta P$, we obtain
\beq
\delta {\cal F} = \int dV \  \left[ - \gamma \nabla^2 P - 
a |P| + 
b |P|^3  \right ](\delta P)  + 
\int dS  \left [\gamma (\hat n \cdot \vec\nabla P) + \alpha (T - T^*) |P|\right] (\delta P).
\label{variationalF}
\enq
In bulk systems the surface integral in $(\ref{variationalF})$ is often
neglected, either due to periodic boundary conditions or due to the
absence of variation in $P$ on very long length-scales.  However this
approach cannot be justified for finite-size systems.

For simplicity, let us restrict our attention
here to the case where the polarization is only $z$-dependent,  
the surface defined as $z=0$,  and the sample occupies the space $z>0$.  
Then the stationary 
condition for the total free energy to a small variation $\delta P(z)$ 
leads to the second-order differential equation
\beq
- \gamma \frac{d^2 P}{dz^2} - a P + b P^3 = 0
\label{eq1}
\enq
together with the boundary condition
\beq
\gamma \frac{dP}{dz} + \alpha (T - T^*) P = 0 \vert_{z \rightarrow 0^+}
\label{bcsurface}
\enq
at the surface.    
We can rewrite the boundary condition as
\beq
\frac{dP}{dz} = - \frac{P}{\delta} \vert_{z \rightarrow 0^+}
\label{deltadef}
\enq
which results in
\beq
P = P(0) e^{-\frac{z}{\delta}} \vert_{z \rightarrow 0^+}
\label{Psurf}
\enq
with 
\beq
\delta = \frac{\gamma}{\alpha (T - T^*)}
\label{delta}
\enq
where we note that the dimensions of $\gamma$ and 
$\alpha (T - T^*)$ are $\frac{E}{P^2 L}$ and $\frac{E}{P^2 L^2}$
respectively so 
that $[\delta] = L$ and $\delta$ is the so-called extrapolation length; here
$P(0)$, the polarization at the surface, must be determined by an electrical
boundary condition.
We remark that 
we must obtain
\beq
|\delta | > \xi_0
\label{scondition}
\enq
in order for this calculation to be self-consistent, where $\xi_0$ is the
length-scale over which the polarization has spatial variation in the bulk.
We see that $\delta$ is temperature-dependent, and its
sign is determined by
the relative values of $T^*$ and $T$.  
For example if, as we have assumed above, $T < T_0$,  
and $T_0 < T^*$, then $(T - T^*) < 0$ and $\delta < 0$.  In this case, 
the surface becomes
ferroelectric at a temperature higher than that in the bulk, so
the polarization at the surface increases as indicated in $(\ref{Psurf})$.  
By contrast, it will
decrease at the boundary if $T_0 > T^*$.
This approach has been generalized to ferroelectric films
\cite{Tilley84,Ong01,Tilley96} with the appropriate 
boundary conditions at
the surfaces/interfaces. The solution of (\ref{eq1}) and the film analogues 
of (\ref{Psurf}) leads to a thickness-dependent critical temperature
that is qualitatively similar to that 
observed \cite{Tilley84,Ong01,Tilley96}.
It can also be extended to the case where the bulk material has
a first-order transition \cite{Scott88}; now higher-order terms in the
Landau-Ginzburg free energy must be retained. Another generalization
is to superlattices with alternating films of two different 
materials \cite{Tilley96,Tilley88}. 

The scheme that we have just outlined assumes implicitly that the free
energy of interest can be clearly separated into two distinct
parts, ${\cal F}_{interior}$
and ${\cal F}_{surface}$, that have the similar algebraic structures but
different transition temperatures. 
More specifically,
there must be a distinct separation 
of length-scales associated
with the surface and the interior; this requirement translates 
into the inequality
\beq
\xi_0 < |\delta| < L
\label{lengthscales}
\enq
where $L >> 1$ is the lateral dimension of the sample.
To our knowledge, there is 
no direct experimental confirmation of the extrapolation length, though 
it may be indirectly accessible via x-ray diffraction \cite{Ishikawa99}
and far-infrared. 
We note
that there is some suggestion \cite{Basceri97} that the Landau coefficients
for thin films may have some thickness-dependence that is not included
in this standard treatment.  Another complementary approach is to assume
a spatially varying transition temperature in the finite-size 
ferroelectric \cite{Li02}. This might be a productive way of incorporating
known imperfections into the phenomenological theory.  Finally we remind
the reader that the Landau-Ginzburg approach assumes that the free energy can
be expressed as a polynomial expansion of the average polarization;
as finite-size effects become increasingly important such coarse-graining
may no longer be possible.  The possible
breakdown of this free energy ansatz should be
checked explicitly with first-principles
methods for a variety of sizes and geometries.

\subsection{Depolarization Effects}

So far we have discussed finite-size planar ferroelectrics where
the polarization has been parallel to the surfaces of interest.  Now
we switch to a geometry
where the polarization is normal to the boundary, as displayed in 
\figref{fig:depol}.  Physically we expect
a build-up of free surface charge which, if uncompensated,
results in a depolarizing electric field $E_{d}$.                    
It is then energetically difficult
for the sample to sustain its uniform polarization, and
phases with ferroelectric domains often result. Since, as we shall see shortly,
the depolarization field increases with decreasing film 
thickness \cite{Batra72,Mehta73,Wurfel73}, 
its importance increases
with reduced size and it must be incorporated in the phenomenological
description of such finite-size ferroelectric systems.

\begin{figure}
     \centerline{\includegraphics[height=5in]{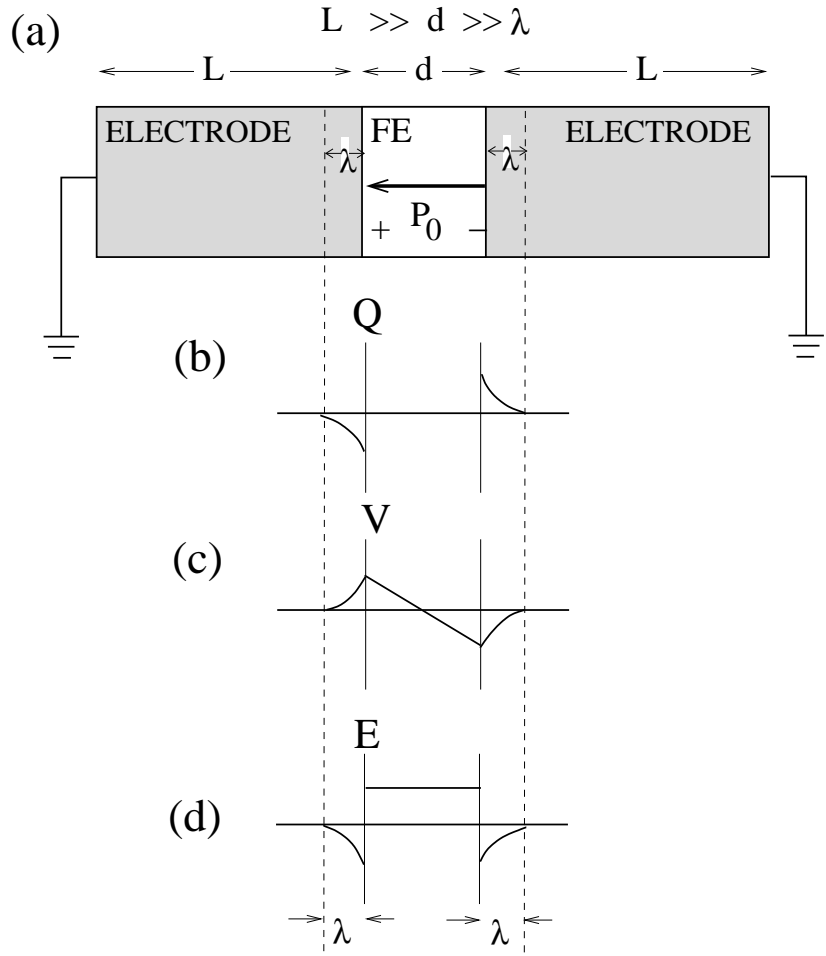}}
     \caption{\label{fig:depol} {\bf a)} Schematic of short-circuited ferroelectric capacitor
with $L >> d >> \lambda$ and accompanying plots of the spatial distribution of
{\bf b)}  the charge $Q$, {\bf c)} the potential $V$ and {\bf d)}  the depolarization field $E$.  
Note that the charge
distribution in {\bf b)} would simply be $\delta$-functions at the film-electrode interfaces for perfect
electrodes; then $V$ and $E$ would simply be constant and zero. }
     \end{figure}

In practice these depolarization effects can be significantly reduced
by the presence of metal electrodes that provide charge compensation
at the ferroelectric boundaries.  We now develop an expression for
the Landau-Ginzburg free energy of a short-circuited ferroelectric capacitor
that consists of a ferroelectric film bracketed by two metal electrodes
(see \figref{fig:depol}) 
For pedagogical simplicity, we'll consider the
short-circuited case ($V_{ext} = 0$) where
the length-scale of the electrodes is significantly larger than that
of the film 
($L  >> d$; $L \rightarrow \infty$),
which is in turn greater than the screening length
($\lambda$) in the electrodes; succinctly our operating condition
is then $L  >> d >> \lambda$ and a schematic of this situation is
displayed in \figref{fig:depol}.  We will also assume that  
the polarization gradient
is negligible at the ferroelectric-electrode interfaces, namely that 
$\frac{dP}{dz} = 0\vert_{z=\pm \frac{d}{2}}$, which implies that
the inverse extrapolation length in (\ref{deltadef}) is zero (e.g.
$\delta \rightarrow \infty$).  .
We note that calculations in this simplified
parameter regime yield results that are compatible with 
experiment \cite{Wurfel73,Dawber03}.   The interested reader 
is referred to more complex treatments of the depolarization field 
for other parameter specifications \cite{Tilley93}.

Let us consider the zero-potential (short-circuited) ferroelectric
capacitor shown schematically in \figref{fig:depol} a).  
The spontaneous polarization
is displayed.  In an idealized system with perfect metal plates,
there would be complete charge compensation at the electrode-ferroelectric
interface and thus no depolarization field in the film.  
However in realistic electrodes, screening charge resides within a small
but nonzero distance from the ferroelectric boundary (cf. \figref{fig:depol} b).  For
a short-circuited ferroelectric capacitor, this spatial charge distribution
leads to an associated voltage drop in the electrodes and a 
compensating
depolarization potential across the film (cf. \figref{fig:depol} c).  There is therefore 
an accompanying
depolarization field which, in the case of uniform polarization, will simply
scale inversely with film thickness (\figref{fig:depol}).  Intuitively we 
expect the 
presence of this depolarization field to suppress ferroelectricity since it
is antiparallel to the polarization.

More formally we follow previous treatments \cite{Tilley93} in the specific 
limit $L >> d >> \lambda$ 
to derive the total Landau free energy of the ferroelectric capacitor;
this free energy has two contributions
\beq
F = F_e + F_f
\label{F}
\enq
where $F_e$ and $F_f$ are associated with the electrodes and the film 
respectively.  We assume that the charge density at each ferroelectric-electrode
interface takes the simple Thomas-Fermi form 
$\rho(z)= \frac{Q}{z} e^{-\frac{z}{\lambda}}$
where $\lambda$ is a screening length \cite{Kittel76} .
Then, solving for the free energy associated with the field inside each
electrode, we find that
\beq 
F_e = \frac{\lambda Q^2}{2 \epsilon_0 \epsilon_e}
\label{Fe}
\enq
where $\epsilon_e$ and $\epsilon_0$ are the permittivities
of the electrodes and of free space respectively.  We note that 
the choice of $\epsilon_e$, a quantity usually defined in the long-wavelength 
limit \cite{Ashcroft76},
is tricky since it enters this treatment
via a boundary condition at the ferroelectric-electrode interface,
and we refer the interested reader to two distinct approaches to
this issue \cite{Dawber03,Tilley93} where comparisons with
experiment are made.

Next we turn to the free energy in the film, which has the form
\beq
F_f = \int_{-d/2}^{d/2} 
\left\{ {\cal F}_P - \frac{1}{2} E_d (z) P(z) \right\}
dz
\label{Ff}
\enq
where $E_d (z)$ is the depolarization field and the second term in (\ref{Ff})
above represents its self-energy (and hence the factor of $\frac{1}{2}$ 
\cite{Kretschmer79}).  Using the fact that there is
no free charge in the ferroelectric, we write the depolarization field
in the form
$E_d (z) = E_0 - \frac{P(z)}{\epsilon_0}$
where by continuity at the interface 
$E_0 = E(\pm \frac{d}{2}) = \frac{Q}{\epsilon_0}$.  
Now we need to find $E_0$ in terms of $P(z)$ since we do not know $Q$. 
Use of the short-circuit
condition leads to the expression 
\beq
E_0 = \frac{1}{\epsilon_0 (2 \lambda + d)} 
\int_{-\frac{d}{2}}^{+\frac{d}{2}} P(z) dz
\label{E0}
\enq
which can then be inputted into the expressions $(\ref{Fe})$ and $(\ref{Ff})$
for $F_e$ and $F_f$ respectively.  We note that for $P$ uniform and
$\lambda << d$, the depolarization field takes the form
\beq
E_d = E_0 - \frac{P}{\epsilon_0} = \frac{P}{\epsilon_0} 
\left\{ \frac{1}{1 + \frac{2\lambda}{d}} - 1 \right\} 
\sim \frac{2P\lambda}{\epsilon_0d}
\label{Ed2}
\enq
which varies inversely with the film thickness ($d$) and 
vanishes in the case of perfect electrodes ($\lambda = 0$); we note
that a similar result for $E_d$ is found by introducing an air gap of 
thickness $\lambda$ between the film
and perfect electrodes \cite{Dawber03,Bratkovsky01} thereby displacing
the compensating charge from the ferroelectric interface.

The Landau free energy for the ferroelectric
capacitor then is 
\beq
F =  \left(\int_{-\frac{d}{2}}^{+\frac{d}{2}} {\cal F}_P \ dz\right) + \Delta F
\label{DelF}
\enq
where
\beq
\Delta F = \int_{-\frac{d}{2}}^{+\frac{d}{2}}  
\frac{P^2(z)}{2 \epsilon_0} dz  
- \frac{1}{2\epsilon_0 d (1 + \frac{2\lambda}{d})}
\left\{ 1 - \frac{\lambda}{\epsilon_e d ( 1 + \frac{2\lambda}{d})}\right\} 
\left[\int_{-\frac{d}{2}}^{+\frac{d}{2}} P(z) dz \right]^2
\label{Ftotal}
\enq
where we reiterate that this free energy is for the limits
$L >> d >> \lambda$ where 
$L \rightarrow \infty$ and $\delta \rightarrow \infty$.
We see from $(\ref{Ftotal})$ that if $P$ is uniform and 
$\lambda = 0$, then $\Delta F = 0$. However if $P = P(z)$,
$\Delta F$ is finite even for perfect electrodes \cite{Kretschmer79}.
For uniform $P$ and finite $\lambda$ where $\frac{2\lambda}{d} << 1$,
$(\ref{Ftotal})$ yields
\beq
\frac {\Delta F}{d}  \sim \frac{P^2 }{2\epsilon_0} 
\left[ \frac{\lambda}{\epsilon_e d} \right]
\label{DeltaFP}
\enq
which contributes to the coefficent of the quadratic term
in the overall free energy density.  As a result, the
expression
for the film's transition temperature $T^*$ is
\beq
\frac{1}{2} a_0 (T^* - T_0) + \frac{\lambda}{2 \epsilon_e\epsilon_o d } = 0
\enq
which leads to 
\beq
T^* = T_0  - \frac{\lambda}{\epsilon_0 \epsilon_e  d a_0}
\label{T*}
\enq
that is shifted from its bulk value ($T_0$) by a term due to the
depolarization field; we note that when $2\lambda \sim d$,
a more complicated expression results \cite{Tilley92}
that suggests  
that
there exists a minimum thickness $d_{min}$
below which ferroelectricity is unstable 
\cite{Batra72,Mehta73,Tilley96}.
For a given material, characterized by $a_0$ and $T_0$, this
minimum thickness can be tuned by the choice of electrodes 
\cite{Ghosez00,Junquera03,Dawber03} whose relevant characteristics
enter $d_{min}$ via
$\lambda$ and $\epsilon_e$.  However recent 
experiments \cite{Fong04,Fong05,Kim05} 
indicate the presence
of ferroelectricity below this minimum cutoff, which 
may be consistent with theoretical expectations 
for a polydomain ferroelectric state \cite{Bratkovsky06}.
Naturally the depolarization effects will be significantly
reduced if the ferroelectric film is treated
as a semiconductor \cite{Watanabe98} rather than as an insulator.
Before closing this section, we note that recently the dielectric 
properties of 
ferroelectric-paraelectric multilayers
have also been studied within this type of thermodynamic approach, and
it has been found that there exists a critical paraelectric film 
thickness such that ferroelectricity is completely suppressed due
to interlayer electrostatic interactions \cite{Roytburd05}. 
We emphasize that the Landau-Ginzburg approach presented to this
point has assumed a single-domain (homogeneous) ferroelectric
phase, and its generalization to include inhomogeneous effects
will be discussed later in this chapter.

\subsection{Misfit Epitaxial Strain}

The thermodynamic behavior of a ferroelectric film is also
sensitive to mechanical boundary conditions due to the strong
coupling between the polarization and the strain.  Application of
pressure, both hydrostatic \cite{Merz50} and biaxial \cite{Forsbergh54},
is known to affect the transition temperature of bulk perovskite
ferroelectrics.  More recently homogeneous epitaxial strain associated with
substrate-film lattice mismatch has been included in a Landau 
free energy \cite{Pertsev98};
this approach results in strain-induced shifts in the transition
temperature and the spontaneous polarization whose signs depends
on details of specific elastic compliances and electrostrictive
constants. Furthermore, theoretical studies 
\cite{Pertsev98,Pertsev99,Ban02,Dieguez04} of 
temperature-misfit strain phase diagrams indicate that there are observable
``epitaxial phases'' that would not be stable in the bulk.  
Indeed the effects of homogeneous misfit strain have been characterized
and controlled to such a degree that $BaTiO_3$ films
have been strain-engineered so that their observed spontaneous polarization
is more than twice that in the bulk \cite{Choi04}.  Furthermore room-temperature
ferrolelectricity has been observed in strained $SrTiO_3$ films \cite{Haeni04},
where we note that bulk $SrTiO_3$ remains paraelectric down to the lowest 
observable temperatures \cite{Muller79}.

Let us be more specific.  In the phenomenological treatment
of epitaxially strained ferroelectric films grown on thick 
substrates \cite{Pertsev98,Pertsev99},
the finite-size ferroelectric is approached as a bulk material 
with homogeneous elastic terms constrained to match the substrate 
lattice conditions. 
Here it is implicit that 
\beq
L >> d_c > d > \xi_0
\label{dL}
\enq
where $L$ and $d$ are the substrate and
the film thicknesses respectively, $\xi_0$ is the correlation length and
$d_c$ is the critical film thickness above which elastic defects
appear; $d_c$ varies roughly as the inverse of
the lattice mismatch and is usually approximately 
$10 nm $ \cite{Choi04}.
With this set of modelling assumptions, the in-plane
strains, $\eta_1$, $\eta_2$ and $\eta_6$ (here we follow the conventional Voigt
notation) are constant throughout the film thickness and are completely
controlled by the substrate-film lattice mismatch.  
The associated stresses
$\sigma_1$, $\sigma_2$ and $\sigma_6$ are finite but not fixed.  By contrast,
\beq
\sigma_3 = \sigma_4 = \sigma_5 = 0
\label{bc1}
\enq
since there are no
tractions acting on the top film surface.  For the special case of a
(001) ferroelectric film grown on a cubic paraelectric substrate
\beq
\eta_1 = \eta_2 = \bar{\eta} = \frac{a - a_0}{a}
\label{bc2}
\enq
where $a$ and $a_0$
are the constrained and the free film lattice constants respectively. 
For this simple
case of cubic symmetry $\eta_6 = 0$ since the angle between the two lattice
vectors remains unchanged ($\theta = \frac{\pi}{2}$).
$(\ref{bc1})$ and $(\ref{bc2})$ represent mixed mechanical boundary
conditions associated with two-dimensional 
clamping, and the standard elastic free energy 
${\cal F}(P,\sigma)$ cannot
be used to find the equilibrium properties of these 
systems \cite{Forsbergh54,Pertsev98}.  Instead
a Legendre transformation to a modified thermodynamic potential
\beq
{\tilde{\cal F}} = {\cal F} + \eta_1 \sigma_1 + \eta_2 \sigma_2 + \eta_6 \sigma_6 
\label{modF}
\enq
must be performed in order to study the equilibrium properties of
this constrained film.

For pedagogical simplicity, we consider a uniaxial ferroelectric
where $P$ is the polarization in the z-direction.  The free energy,
with condition (\ref{bc1}),
of a cubic ferroelectric is \cite{Haun87}
\beq
{\cal F}  = {\cal F}_P - \frac{1}{2} s_{11} (\sigma_1^2 + \sigma_2^2) - 
Q_{12} \{ (\sigma_1 + \sigma_2) P^2 \} - s_{12} \sigma_1 \sigma_2
- \frac{1}{2} s_{44} \sigma_6^2
\label{Felas}
\enq
where $Q_{ij}$ and $s_{ij}$ are the electrostrictive constants and the
elastic compliances at constant polarization respectively.
Using $\frac{\partial {\cal F}}{\partial \sigma_i} = - \eta_i$, and solving
for $\sigma_1 = \sigma_2 = \bar{\sigma}$ (in this special case $\sigma_6 = 0$),
we find that
\beq
\tilde{{\cal F}} = \frac{\bar{\eta}^2}{s_{11} + s_{12}} +
\frac{1}{2} \tilde{a} P^2
+ \frac{1}{4} \tilde{b} P^4 + \frac{1}{6} c P^6
\label{tilF1}
\enq
where
\beq
\tilde{a} = a - \frac{4 \bar{\eta}Q_{12}}{s_{11} + s_{12}}
\label{atil}
\enq
and 
\beq
\tilde{b} = b + \frac{4 Q_{12}^2}{s_{11} + s_{12}}
\label{btil}
\enq
so that the coefficients of both the quadratic and the quartic polarization
terms in the Landau free energy 
are renormalized.  This has important implications for the thermodynamic
properties of the thin film; more specifically it means that its 
transition temperature is shifted from the bulk value ($T_0$)
\beq
T^* = T_0  + \frac{4 \bar{\eta} Q_{12}}{a(s_{11} + s_{12})}
\label{Tcepitax}
\enq
as is its spontaneous polarization (please see the discussion preceding
$(\ref{Ps})$) due to the in-plane straining of the film by the compressive
substrate.  The signs of these shifts will be determined by the relative
signs and magnitudes of  
$Q_{12}$, $s_{11}$ and $s_{12}$.  We note that such misfit epitaxial
strain can change the nature of the transition from first- to second-order
($b < 0$ but $\tilde{b} > 0$)  For ferroelectric films with multicomponent
polarization and with different orientations, this approach yields a temperature-strain phase diagram
with equilibrium phases that are not stable in a mechanically-free bulk
sample \cite{Pertsev98,Pertsev99,Ban02,Dieguez04,Tagantsev02}. 
The resulting low-temperature phase diagrams are very 
sensitive to the Landau coefficients, particularly for increasing
misfit strain; here the phenomenological approach is nicely complemented
by first-principles studies that can resolve uncertainties associated
with these input parameters \cite{Dieguez04}.
This combined phenomenological-first principles approach has been
very successful in explaining the phase behavior of epitaxially
strained films and we refer the interested reader to a sample of recent 
experimental papers in this field \cite{Choi04,Lu05,Soh06,Rios06,Choi06}.
For ultrathin ferroelectric films with biaxial compressive strain at
the substrate interface, the competition between elastic and surface
effects has also been considered \cite{Zembilgotov02}.
In particular we note that if we do include a gradient (Ginzburg)
term to the free energy then the epitaxial strain leads to a modified
correlation length 
\beq
\tilde{\xi} = \sqrt{\frac{\gamma}{\tilde a}}
\label{xistar}
\enq
where $\xi = \sqrt{\frac{\gamma}{a}}$ is that associated with 
the bulk material that is mechanically free \cite{Zembilgotov02}.
Similar mechanical boundary conditions have also been
studied in epitaxially-strained superlattices \cite{Neaton03}.
We note that recently a combined numerical-thermodynamic approach 
resulted in generalized temperature-strain phase diagrams for thin
ferroelectric films that
include multidomain phases \cite{Li03} inaccessible by the
approaches described here that assume homogenous polarization. 

\subsection{Inhomogeneous Effects}

In the previous discussion, we assume that the energy cost 
of uniformly straining the film to lattice-match the substrate is 
significantly less than that associated with other strain relaxation 
mechanisms; these include the creation of elastic dislocations, 
polydomain formation and even multiphase coexistence which are 
all inhomogeneous in nature.  We expect elastic homogeneity
for films thinner than a critical thickess ($d < d_c$), where
$d_c$ scales (roughly) inversely with the lattice mismatch \cite{Choi04}. 
Typically such coherent epitaxially strained films have thicknesses 
$d \sim 10 \ nm$, though film growth parameters can be tuned to achieve 
$d \sim 50 \ nm$ by kinetically suppressing other elastic relaxation 
processes \cite{Choi04}. However the physics of thicker ferroelectric 
films, particularly those used in current applications  ($d \sim 120 \ nm$) 
\cite{Dawber05}, does include inhomogenous effects and next we review 
aspects of this topic within a phenomenological framework.

The misfit epitaxial strain decreases roughly 
exponentially \cite{Matthews74,Kim99,Sinnamon02}
with distance ($\eta(z) \sim \eta_0 e^{-\frac{1}{\lambda}}$)
from the film-substrate interface with a ``strain depth'' of approximately
$\lambda \sim 300 \ nm$ and thus is negligible for $ d >> \lambda$.
In films of thickness $d_c < d < \lambda$, elastic defects often form
to accomodate film-substrate lattice mismatch.  While permitting
the film's lattice constants to relax, these defects do generate inhomogenous
strains which couple to the film's polarization and therefore
affect its ferroelectric properties.  We note that compositional and thermal 
gradients can also produce such inhomogeneities \cite{Ban03}. Here 
we will focus
on such elastic effects. Indeed the epitaxial phenomenological
treatment can be generalized to include isotropic inhomogenous
strain due to such lattice defects \cite{Balzar04}; this 
results in further contributions to the
coefficients of both $P^2$ and $P^4$ in the free energy, 
and thus to a defect-induced change in the transition temperature  
qualitatively consistent with 
experiment \cite{Balzar04}.

The broadening of the dielectric peak in ferroelectric films with decreasing
thickness has been observed by 
several groups \cite{Sinnamon02,Shaw99,Parker02,Lookman04}.
In bulk systems, such behavior is often attributed to disorder and/or
low-dimensional effects \cite{Cardy96} but these explanations
are not appropriate to ferroelectric films due to the underlying
long-range interactions.
Indeed we expect these dipolar systems to be more sensitive to
electromechanical boundary conditions than to local fluctuations.  This
qualitative idea has been 
confirmed by measurements on ferroelectric
free-standing lamellae of sharp dielectric peaks,
strongly indicating that it is due to substrate-related 
effects most probably associated with interfacial lattice 
mismatch \cite{Saad04}. 

Recent experiments on the flexoelectric coupling between strain gradients 
and the polarization indicate that these effects could be important in 
thin films \cite{Ma01,Gruverman03}.  Theoretically it has been shown that this 
flexoelectric effect is enhanced in high-permittivity materials
such as ferroelectrics \cite{Tagantsev86} where the coupling is maximized 
near the dielectric peak.  This flexoelectric coupling has been incorporated 
into a Landau-Ginzburg framework with the goal of studying its
effect on the thermodynamic properties of epitaxially strained ferroelectric 
films \cite{Catalan04}.  The strain gradient
contribution, $\Delta_{FC}  = \gamma \frac{d\eta(z)}{dz}$, couples to 
the polarization
in the free energy, acting as an effective field, 
so that the equation for the spontaneous polarization
\beq
{\tilde a} P + {\tilde b} P^3 = \Delta_{FC} \vert_{P = P_0}
\label{P_FC}
\enq
can only have finite solutions, even in the absence of an external
electric field ($E=0$); here ${\tilde a}$ and ${\tilde b}$ are the Landau 
coefficients that have been renormalized by epitaxial strain.  
As a result the inverse dielectric susceptibility
\beq
\chi^{-1} = 3 {\tilde b} P_0^2 + {\tilde a}
\label{chi}
\enq 
can never be zero, so the standard singularity in $\chi$ is
replaced by a broadened peak.  Similarly the temperature scale
associated with the onset of reversible polarization is distinct
from that of the maximum of $\chi$.  This approach also yields
a polarization that increases with decreasing film thickness
despite the associated reduction and broadening of the dielectric
peak; these seemingly contradictory features are consistent with 
known experiment \cite{Catalan04}.

We have just seen that the presence of strain gradients via the flexoelectric
effect leads to the broadening of the dielectric peak in ferroelectric
films.  This phenomena can be generalized to consider other effective
fields that couple linearly to the polarization \cite{Bratkovsky05}; 
possible origins of this effective field include 
asymmetric electrodes, compositional and temperature gradients as
well as the stress profiles already considered.  Qualitatively the
effects on the dielectric response will be the same, though naturally
quantitative differences will depend on the physical origin of the
effective field.

Inhomogeneities, particularly strain and temperature gradients, can also lead to multidomain
formation \cite{Bratkovsky00} as an overall energy reduction mechanism; this behavior has been 
observed in relatively thick ($d \sim 500 \ nm$) films \cite{Roytburd01}. The competition
between different strain relaxation processes including domain formation and misfit dislocatio has been studied theoretically \cite{Speck94,Slutsker99}. 
Within the framework of a Landau-Ginzburg theory,
the development of domain structure has been 
addressed \cite{Bratkovsky06,Wang95,Bratkovsky00}; 
qualitatively
this is done by seeking polarization solutions of the form
\beq
P = P_0 + \delta P_k(z) e^{ikx}
\label{Pk}
\enq
where $\delta P_k(z) << P_0$.  Such a treatment in nearly cubic ferroelectrics
indicates that a multidomain 
state may well be stable down to atomic film thicknesses, length-scales
well below the previous estimates that only described
single-domain ferroelectricity \cite{Bratkovsky06}.The time evolution of
the polarization and therefore the resulting domain structures
can also be obtained by numerical studies of the time-dependent Landau-Ginzburg equations\cite{Chen98};
such phase-field studies predict specific domain morphologies \cite{Li05a,Li05b,Li06a,Wang04} for epitaxial
thin films of a variety of materials including $PbZr_{1-x}Ti_xO_3$, $SrTiO_3$ and $BaTiO_3$.
We note that
inhomogeneities in multicomponent ferroelectrics may also lead to multiphase
coexistence \cite{Slutsker99,Semenovskaya97,Roytburd02, Li06b}, particularly in the region of a first-order phase transition.  

\section{Summary and (Some) Open Questions}

At this point we have discussed the main features associated with
ferroelectrics that have been addressed within a phenomenological 
Landau-Ginzburg theory.  Due to their underlying long-range interactions,
we have learned that bulk ferroelectrics are well described within this
framework; for the same reason, their 
finite-size counterparts are very sensitive to electromechanical 
boundary conditions including surface
charge and misfit epitaxial strain.  Inhomogeneous effects, particularly
strain gradients, have also been discussed.  We have now worked our way
to the present and naturally there remain a number of challenges that
can be addressed within this framework.  Let us discuss a few of these:
\\
\noindent 
{\bf (i) Strain-gradients and finite-size effects in novel nanogeometries }
  
Three-dimensional ferroelectric nanogeometries are
being explored as key components in competitive high-density data storage
devices \cite{Arimoto04}.  The complexity of their topologies and their 
boundary conditions
tends to favor inhomogenous polarization and strain 
configurations (e.g. \cite{Scott05,Fu03}) which remain
largely unstudied, particularly from a phenomenological standpoint.
For example, the energetics associated with the formation of misfit
dislocations in ferroelectrics with curved topologies is known to
be different than those in planar structures \cite{Gutkin00}, and the
associated strain gradients could lead to complex three-dimensional
polarization patterns.
\\
\noindent
{\bf (ii) Multiferroics}
 
There exist a number of materials that display both long-range ferroelectricity
and incommensurate magnetic order \cite{Khomskii06} and we refer
the interested reader to the chapter by N. Spaldin on this topic.  
Landau 
treatments \cite{Lawes05,Mostovoy06} for  this class of materials have 
provided symmetry-based arguments for
the type of underlying interactions that must be present at the microscopic 
level; furthermore the couplings between ferroelectric and magnetic order parameters can be studied in nanostructured electromagnets \cite{Zheng04}. Detailed interplay between microscopic and phenomenological models 
should provide guidance on material parameters necessary to enhance the 
magnetoelectric couplings, particularly at room temperature for 
device applications.
\\
\noindent
{\bf (iii) Dynamics}

The dynamical dielectric response of ferroelectrics should be accessible
via time-dependent Landau-Ginzburg theory, similar to studies performed with this
method in ferroelastics (e.g.\cite{Shenoy99}).  
More generally a detailed phenomenological study of domain 
nucleation \cite{Ganpule00}, domain motion 
(and domain interlocking \cite{Krishnan02}) could be particularly useful
for the characterization of switching properties of 
thin-films \cite{Fousek03}.
Also glassy behavior has been observed in a number of ferroelectric
relaxor materials \cite{Colla00}, and a phenomenological model of 
these relaxational dynamics should be feasible.
\\
\noindent
{\bf (iv) Possible Breakdown of Landau-Ginzburg theory}

Landau-Ginzburg theory may well break down in ultrasmall ferroelectric 
structures where the necessary averaging of the order parameter (i.e. the 
polarization) is no longer possible.  Also it remains a challenge to see
how this highly local approach can be used to described nonlocal elastic and 
Coulomb effects, particularly vis a vis domain formation and energetics.
\\
\noindent
{\bf (v) Quantum Critical Fluctuations }

The paralelectric-ferroelectric transition in the limit $T \rightarrow 0$
may well be a textbook candidate \cite{Grupp97,Coleman05,Scott06b} for quantum critical 
behavior \cite{Sachdev99}. Though some theoretical 
predictions \cite{Roussev03} exist, logarithmic corrections to 
mean-field (Landau) scaling should be both theoretically and experimentally 
accessible and have yet to be explored.  As in the classical 
case \cite{Larkin68},
ferroelectric materials may well provide the setting for the first 
detailed study
of criticality at a quantum phase transition.
\\
\\
These are only some of the many outstanding challenges that the Landau-Ginzburg
theory of ferroelectrics still has to address.   
Most generally, it can provide a solid 
bridge between the physics on atomistic scales and macroscopic measurable
quantities, 
particularly in finite-size ferroelectrics where electromechanical boundary 
conditions are crucial and therefore the physics spans many length-scales.
In closing, we invite the (hopefully still) curious reader to partake in 
these and other adventures
with this phenomenological approach to ferroelectric materials.

\section{Acknowledgements}

We are grateful to M. Dawber, C. Fennie, L. Palova, K.M. Rabe and J.F. Scott
for their careful readings
of this text, and more generally for numerous discussions 
about the topics therein.  
We would also like to thank O. Dieguez and P. Coleman
for clarifying aspects of misfit epitaxial strain 
and surface superconductivity
respectively for us. 
PC acknowleges support from NSF NIRT DMR-0210575 and the Aspen
Center for Physics for hospitality.
PBL acknowledges support from the Engineering and Physical Sciences Research Council, and the hospitality of the National High Magnetic Field Laboratory.


\begin{thebibliography}{99}

\bibitem{Landau59} L.D. Landau and E.M. Lifshitz, {\sl Statistical Physics},
(Pergamon Press, Oxford 1959).

\bibitem{Toledano87} J.-C. Toledano and P. Toledano, {\sl Landau Theory of Phase Transitions} (World Scientific, Singapore, 1987).

\bibitem{Fatuzzo67} E. Fatuzzo and W.M. Mertz, {\sl Ferroelectricity} (North-Holland, Amsterdam, 1967).

\bibitem{Lines77} M.E. Lines and A.M. Glass, {\sl Principles and Applications 
of Ferroelectrics and Related Materials}, (Oxford University Press, Oxford, 1977).

\bibitem{Jona93} F. Jona and G. Shirane, {\sl Ferroelectric Crystals}, (Dover, NY, 1993).

\bibitem{Strukov98} B.A. Strukov and A.P.Levanyuk, {\sl Ferroelectric 
Phenomena in Crystals}, (Springer, Berlin, 1998).

\bibitem{Scott00} J.F. Scott, {\sl Ferroelectric Memories}, Springer Series in Advanced Microelectronics, Vol. 3 (Springer, Berlin, 2000).

\bibitem{Landau37} L.D. Landau, {\sl Phys. Z. Sowjun.} {\bf 11}, 26 , 545 (1937); Zh. Eksp. Teor. Fiz. {\bf 7} 627 (1937); English translation in {\sl Collected Papers of L.D. Landau}, ed. D. ter Haar (Pergamon Press, Oxford, 1965).

\bibitem{Mueller40} H. Mueller, {\sl Phys Rev} {\bf 57}, 829 (1940); Phys. Rev. {\bf 58}, 565.

\bibitem{Devonshire49} A.F.Devonshire, {\sl Phil. Mag.} {\bf 40}, 1040 (1949); {\sl ibid} {\bf 42}, 1065 (1951); {\sl Adv. Phys.} {\bf 3}, 85 (1954).

\bibitem{Ginzburg45} V.L. Ginzburg, {\sl Zh. Eksp. Teor. Fiz.} {\bf 15}, 739 (1945); {\sl ibid} {\bf 19}, 36 (1949).

\bibitem{Levanyuk59} A.P. Levanyuk, {\sl Sov. Phys. JETP} {\bf 36}, 571 (1959).

\bibitem{Ginzburg60} V.L. Ginzburg, {\sl Fiz. tverd. Tela} {\bf 2} 2031 (1960); English
translation in {\sl Sov. Phys.- Solid State} {\bf 2}, 1824 (1960).

\bibitem{Born54} M. Born and K. Huang, {\sl Dynamical Theory of Crystal Lattices}, (Clarendon Press, Oxford, 1954)

\bibitem{Larkin68} A.I. Larkin and D.E. Khmel'nitskii, {\sl Zh. eksp. teor. Fiz.} 
{\bf 55}, 2345 (1968); {\sl ibid} {\bf 56}, 2087 (1969); English 
translations {\sl Sov. Phys.-JETP} {\bf 28}, 1245 (1969) and 
{\sl ibid} {\bf 29}, 1123 (1969).

\bibitem{Ahlers75} G. Ahlers et al., {\sl Phys. Rev. Lett.} {\bf 34}, 1227 (1975).

\bibitem{Batra72} I.P. Batra and B.D. Silverman, {\sl Solid State Communications} {\bf 11}
291 (1972).

\bibitem{Mehta73} R. Mehta, B. Silverman and J.T. Jacobs, {\sl J. Appl. Phys.} {\bf 44}, 3379 (1973).

\bibitem{Wurfel73} P. Wurfel and I.P. Batra, {\sl Phys. Rev. B} {\bf 8}, 5126 (1973).


\bibitem{Ghosez00} Ph. Ghosez and K.M. Rabe, {\sl Appl. Phys. Lett.} {\bf 76} 
2767 (2000).

\bibitem{Junquera03} J. Junquera and P. Ghosez {\sl Nature} {\bf 422}, 506 (2003).

\bibitem{Dawber03} M. Dawber, P. Chandra, P.B. Littlewood and J.F. Scott, {\sl J. Phys: Cond. Mat.} {\bf 15}, L393 (2003).

\bibitem{Tilley84} D.R. Tilley and B. Zeks, {\sl Solid State Commun.} {\bf 49}
823 (1984). 

\bibitem{Haeni04} J.H. Haeni et al, {\sl Science} {\bf 430} 758 (2004); Y.L. Li et al, {\sl Phys Rev. B} {\bf 73} 184112 (2006).

\bibitem{Li05a} Y.L. Li et al, {\sl J. Appl. Phys.} {\bf 98} 064101 (2005); Y.L. Li et al, {\sl Appl. Phys.} {\bf 88} 072905 (2006).

\bibitem{Li05b} Y.L. Li et al, {\sl J. Appl. Phys.} {\bf 97} 034112 (2005).

\bibitem{Merz50} W. Merz, {\sl Phys. Rev. B} {\bf 77}, 52 (1950).

\bibitem{Forsbergh54} P.W. Forsbergh {\sl Phys. Rev. B} {\bf 93}, 686 (1954). 
%% elastic compatibility
\bibitem{Love44} E. A. H. Love, {\sl A Treatise on the Mathematical Theory of
Elasticity} (Dover, New York, 1944), p. 49.

\bibitem{Kartha95}
S. Kartha, J.A. Krumhansl, J.P. Sethna, and L.K.Wickham, Phys.
Rev. B {\bf 52}, 803 (1995).

\bibitem{Shenoy99}
S. R. Shenoy, T. Lookman, A. Saxena, and A. R. Bishop,
Phys. Rev. B {\bf 60}, R12537 (1999).

\bibitem{Rasmussen01}
K. Ø. Rasmussen, T. Lookman, A. Saxena, A. R. Bishop, and R. C. Albers, Phys. Rev. B {\bf 87}, 55704 (2001).

\bibitem{Lookman03}
 T. Lookman, S.R. Shenoy, K.O. Rasmussen, A. Saxena, and A.R. Bishop,
Phys. Rev. B {\bf 67}, 24114 (2003).
%% end elastic compatibility refs

\bibitem{Fujii71} Y. Fujii and Y. Yamada {\sl J. Phys. Soc. Japan} {\bf 30}, 1676 (1971).

\bibitem{Als-Nielsen77} J. Als-Nielsen and R. Birgeneau {\sl Am. J. of Physics}, {\bf 45}, 554 (1977).

\bibitem{Dima} D. Khmel'nitskii and J.F. Scott, private communication.

\bibitem{White79} e.g. R.M. White and T.H. Geballe, {\sl Long-Range Order in Solids} (Academic Press, New York, 1979).
 
%% new refs on displacive transitions
\bibitem{Cochran60} W. Cochran, {\sl Adv. Phys.} {\bf 9}, 387 (1960).

\bibitem{Cowley62} R.A. Cowley, {\sl Phys. Rev. Lett.} {\bf 9}, 159 (1962).

\bibitem{Bruce81} A.D. Bruce and R.A. Cowley {\sl Structural Phase Transitions }(London: Taylor and Francis), (1981).

\bibitem{Burstein70} E. Burstein, A. Pinczuk, and R.F. Wallace, {\sl Proceedings of the Conference on the Physics of Metals and Narrow Gap Semiconductors}, ed. D.L. Carter and R.T. Bate, (Pergamon, NY), 251-272 (1970).



\bibitem{Littlewood79} P.B. Littlewood and V. Heine, {\sl J. Phys C: Solid St. Phys.}, {\bf 12}, 4431; P.B. Littlewood, {\sl. ibid.}, 4441.



\bibitem{Portengen96} T. Portengen, Th. ¨Ostreich and L. J. Sham, {\sl Phys. Rev. Lett.} {\bf 76},
3384 (1996); {\sl Phys. Rev. B} {\bf 54}, 17452 (1996).

\bibitem{Batista04} C. D. Batista and A. A. Aligia, {\sl Phys. Rev. Lett.} {\bf 92}, 246405 (2004); {\sl Phys. Rev. B} {\bf 71}, 125110 (2005).

\bibitem{Gruner94} G. Gruner, {\sl Density Waves in Solids} (Addison-Wesley, 1994).

\bibitem{Schrieffer64} {\sl Theory of Superconductivity} (Benjamin, Reading, 1964).

\bibitem{deGennes66} P.G. deGennes, {\sl Superconductivity of Metals and 
Alloys} (Benjamin, New York, 1966).

\bibitem{Tinkham75} M. Tinkham, {\sl Introduction to Superconductivity} (McGraw-Hill, New York, 1975).


\bibitem{McMillan77}  W. L. McMillan,
{\sl Phys. Rev. B} {\bf 16}, 643 (1977)

\bibitem{Cowley96} R.A. Cowley, {\sl Physica A} {\bf 232}, 585 (1996).

\bibitem{Efremov03} D.V. Efremov, J van den Brink, and D.I. Khomskii,
{\sl Nature Materials} {\bf 3}, 853 (2004).
\bibitem{Vanderbilt01} D. Vanderbilt and M.H. Cohen, {\sl Phys. Rev. B} {\bf 63}, 094108 (2001).
%% end displacive refs
\bibitem{Noheda99} B. Noheda et al., {\sl Appl. Phys. Lett.}{\bf 74},  
2059 (1999).

\bibitem{Haun89} M.J. Haun et al., {\sl Ferroelectrics} {\bf 99} 13 (1989).

\bibitem{Bellaiche00} L. Bellaiche, A. Garcia and D. Vanderbilt, {\sl Phys. Rev. B} {\bf 52}, 6301 (1995).

\bibitem{Zhong95} W. Zhong, D. Vanderbilt and K.M. Rabe, {\sl Phys. Rev. B} {\bf 52}, 6301 (1995).

\bibitem{Ahn04} C.H. Ahn, K.M. Rabe and J.-M. Triscone, {\sl Science} {\bf 303}, 488 (2004)

\bibitem{Dawber05} M. Dawber, K.M. Rabe and J.F. Scott, {\sl Rev. Mod. Phys.} 
{\bf 77}, 1083 (2005).

\bibitem{Zhong94} W. Zhong, D. Vanderbilt and K.M. Rabe, {\sl Phys. Rev. Lett.} {\bf 73}, 1861 (1994); {\sl Phys. Rev. B} {\bf 52}, 6301 (1995).

\bibitem{Iniguez01} J. Iniguez et al., {\sl Phys. Rev. B} {\bf 63}, 
144103 (2001).

\bibitem{Langer69} J.S. Langer, {\sl Ann. Phys. (NY)} {\bf 54}, 258 (1969).

\bibitem{McWhan85} D.B. McWhan et al., {\sl J. Phys. C} {\bf 18}, L307 (1985);
D.A. Neumann et al., {\sl Phys. Rev. B} {\bf 32}, 1866 (1985).

\bibitem{Littlewood86} P.B. Littlewood and P. Chandra, {\sl Phys. Rev. Lett.} {\bf 57} 2415 (1986); P. Chandra, 
{\sl Phys. Rev. A} {\bf 39} 3672 (1989).

\bibitem{Ginzburg01} V.L. Ginzburg, {\sl Physics-Uspekii} {\bf 44}, 1037 (2001).

\bibitem{Li96} S.P. Li et al. {\sl Phys. Lett.} {\bf 212} 341 (1996); S.P. Li et al. {\sl Jap. J Appl Phys.} {\bf 36} 5169 (1997).

\bibitem{Auciello98} O. Auciello, J.F. Scott and R. Ramesh, {\sl Physics Today} {\bf 51}, 22 - 27 (1998).

\bibitem{Kretschmer79} R. Kretschmer and K. Binder, {\sl Phys. Rev. B} {\bf 20}, 1065 (1979); K. Binder, {\sl Ferroelectrics} {\bf 35}, 99 (1981).

\bibitem{Lubensky75} T.C. Lubensky and M.H. Rubin, {\sl Phys. Rev. B} {\bf 12}, 3885 (1975).

\bibitem{Scott06a} J.F. Scott, {\sl J. Phys. Cond. Mat} {\bf 18} R361 (2006).

\bibitem{Ong01} L-H Ong, J. Osman and D.R. Tilley, {\sl Phys. Rev. B} {\bf 63}, 144109 (2001).

\bibitem{Kaganov72} M.I. Kaganov and A.N. Omelyandchouk, {\sl Zh. Eksp. Teor. Fiz.} {\bf 61}, 1679 (1971) [{\sl Sov. Phys. JETP} {\bf 34}, 895 (1972)].

\bibitem{Binder72} K. Binder and P.C. Hohenberg, {\sl Phys. Rev. B} {\bf 6}, 
3461 (1979).

\bibitem{Tilley96} D.R. Tilley in {\sl Ferroelectric Thin Films:  Synthesis
and Basic Properties} eds. C. Paz de Araujo, J.F. Scott and G.W. Taylor,
(Gordon and Breach, Amsterdam, 1996) pp. 11 - 46.

\bibitem{Scott88} J.F. Scott et al., {\sl Physica B} {\bf 150}, 160 (1988).

\bibitem{Tilley88} D.R. Tilley, {\sl Solid State Comm.} {\bf 65}, 657 (1988).

\bibitem{Ishikawa99} K. Ishikawa and T. Uemori, {\sl Phys. Rev B} {\bf 60} 11841 (1999).

\bibitem{Chew99} K.-H. Chew et al., {\sl Integr. Ferr.} {\bf 23}, 161 (1999).

\bibitem{Basceri97} C. Basceri et al., {\sl J. Appl. Phys.} {\bf 82}, 
2497 (1997).

\bibitem{Li02} T. Li and W. Cao, {\sl Phys. Rev. B} {\bf 66}, 24102 (2002).

\bibitem{Tilley93} D.L. Tilley in {\sl Ferroelectric Ceramics}, eds. N. Setter and E.L. Colla (Birkhauser Verlag, Basel, 1993) pp. 163-184.

\bibitem{Kittel76} C. Kittel, {\sl Introduction to Solid State Physics} (J. Wiley and Sons, New York 1976).

\bibitem{Ashcroft76} N.W. Ashcroft and N.D. Mermin, {\sl Solid State Physics} 
(Holt, Rinehart and Winston, New York, 1976). 

\bibitem{Tilley92} D.L. Tilley, {\sl Ferroelectrics} {\bf 134}, 313 (1992).

\bibitem{Bratkovsky01} A.M. Bratkovsky and A.P. Levanyuk, 
{\sl Phys. Rev. B.} {\bf 63}, 132103 (2001).

\bibitem{Fong04} D.D. Fong et al., {\sl Science} {\bf 304}, 1650 (2004).

\bibitem{Fong05} D.D. Fong et al., {\sl Phys. Rev. B} {\bf 71} 144112 (2005). 

\bibitem{Kim05} Y.S. Kim et al., {\sl Appl. Phys. Lett.} {\bf 86}, 
102907 (2005).

\bibitem{Bratkovsky06} A.M. Bratkovsky and A.P. Levanyuk, cond-mat/0601484.

\bibitem{Watanabe98} Y. Watanabe, {\sl Phys. Rev. B} {\bf 57} 789 (1998).

\bibitem{Roytburd05} A.L. Roytburd et al., {\sl Appl. Phys. Let.} 
{\bf 87} 72909 (2005).

\bibitem{Pertsev98} N.A. Pertsev, A.G. Zembilgotov and A.K. Tagantsev,
{\sl Phys. Rev. Lett.} {\bf 80}, 1988 (1998).

\bibitem{Pertsev99} N.A. Pertsev, A.G. Zembilgotov and A.K. Tagantsev,
{\sl Ferroelectrics} {\bf 223} 79 (1999).

\bibitem{Ban02} Z.-G. Ban and S.P. Alpay, {\sl J.of Appl. Phys} 
{\bf 91} 9288 (2002).

\bibitem{Dieguez04} O. Dieguez et al. {\sl Phys. Rev. B} {\bf 69}, 212101 (2004); O. Dieguez et al., {\sl Phys. Rev. B} {\bf 72} 144101 (2005).

\bibitem{Tagantsev02} A.K. Tagantsev et al., {\sl Phys. Rev. B} 
{\bf 65}, 012104 (2002). 

\bibitem{Choi04} K. Choi et al, {\sl Science} {\bf 306} 1005 (2004).

\bibitem{Muller79} K.A. Muller and H. Burkard, {\sl Phys. Rev. B} {\bf 19} 3593 (1979).

\bibitem{Haun87} M.J. Haun et al., {\sl J. Appl. Phys.}{\bf 62}, 3331 (1987).

\bibitem{Lu05} X.M. Lu et al., {\sl Phys. Rev. B} {\bf 72}, 212103 (2005).

\bibitem{Soh06} A.K. Soh et al., {\sl J. of the Amer. Cer. Soc.} {\bf 89} 652 (2006).

\bibitem{Rios06} S. Rios et al., {\sl J. of Appl. Phys.}{\bf 99}, 
24107 (2006).

\bibitem{Choi06} S.K. Choi et al., {\sl Appl. Phys. Lett.} {\bf 88}, 
52901 (2006). 

\bibitem{Zembilgotov02} A.G. Zembilgotov et al., {\sl J. of Appl. Phys.}
{\bf 91}, 2247 (2002).

\bibitem{Neaton03} J.B. Neaton and K.M. Rabe, {\sl Appl. Phys. Lett.} 
{\bf 82}, 1586 (2003); K. Johnston et al., {\sl Phys. Rev. B} {\bf 71}, 
100103 (2005);
S.M. Nakhmanson et al. {\sl Appl. Phys. Lett.} {\bf 87}, 102906 (2005).

\bibitem{Li03} Y.L. Li et al. {\sl Appl. Phys. Let.} {\bf 83} 1608 (2003);
Y.L. Li and L.Q. Chen {\sl Appl. Phys. Let.} {\bf 88} 72905 (2006).

\bibitem{Matthews74} J.W. Matthews and A.E. Blackeslee, 
{\sl J. Cryst. Growth} {\bf 27}, 118 (1974).

\bibitem{Kim99} H.J. Kim et al., {\sl Appl. Phys. Lett.} {\bf 78} 1724 (2001).

\bibitem{Sinnamon02} L.J. Sinnamon et al., {\sl Appl. Phys. Lett.} 
{\bf 81}, 703 (2002).

\bibitem{Ban03} Z.G. Ban et al., {\sl Phys. Rev. B} {\bf 67}, 184104 (2003).

\bibitem{Balzar04} D. Balzar et al., {\sl Phys. Rev. B} {\bf 70}, 92103 (2004).

\bibitem{Shaw99} T.M. Shaw et al., {\sl Appl. Phys. Lett.} 
{\bf 75}, 2129 (1999).

\bibitem{Parker02} C.B. Parker et al., {\sl Appl. Phys. Lett.}{\bf 81}, 
340 (2002).

\bibitem{Lookman04} A. Lookman et al.,  {\sl J. Appl. Phys.} {\bf 96}, 
555 (2004).

\bibitem{Cardy96} J. Cardy, {\sl Scaling and Renormalization in Statistical
Physics} (Cambridge University Press, Cambridge, 1996).

\bibitem{Saad04} M.M. Saad et al., {\sl J. Phys. C} {\bf 16} L451 (2004).

\bibitem{Ma01} W. Ma and L.E. Cross, {\sl Appl. Phys. Lett.} 
{\bf 79} 4420 (2001); {\sl ibid} {\bf 81} 3440 (2002); 
{\sl ibid} {\bf 82} 3293 (2003); {\sl ibid} {\bf 86} 72905 (2005).

\bibitem{Gruverman03} A. Gruverman et al. {\sl Appl. Phys. Lett.} {\bf 83},
728 (2003).

\bibitem{Tagantsev86} A. Tagantsev, {\sl Phys. Rev. B} {\bf 34} 5883 (1986);
A. Tagantsev {\sl Phase Transitions} {\bf 35} 119 (1991).

\bibitem{Catalan04} G. Catalan et al., {\sl J. Phys. Cond. Mat.} 
{\bf 16}, 2253 (2004); G. Catalan et al., {\sl Phys. Rev. B} {\bf 72} 
020102 (2005).

\bibitem{Bratkovsky05} A.M. Bratkovsky and A.P. Levanyuk, 
{\sl Phys. Rev. Lett.} {\bf 94}, 107601 (2005).

\bibitem{Wang95} Y.G. Wang et al., {\sl Phys. Rev. B} {\bf 51} 5311 (1995).

\bibitem{Bratkovsky00} A.M. Bratkovsky and A.P. Levanyuk, 
{\sl Phys. Rev. Lett.}{\bf 1}, 43177 (2000); A.M. Bratkovsky and A.P. Levanyuk,
{\sl Phys. Rev B} {\bf 66}, 184109 (2002).

\bibitem{Roytburd01} A.L. Roytburd et al., {\sl J. Appl. Phys.} {\bf 89}, 
553 (2001).

\bibitem{Speck94} J.S. Speck and W. Pompe {\sl J. Appl. Phys.} {\bf 7}, 
466 (1994).

\bibitem{Slutsker99} J. Slutsker and A.L. Roytburd {\sl Ferroelectrics} 
{\bf 222} 365 (1999).

\bibitem{Chen98} L.Q. Chen and J. Shen, {\sl Comput. Phys. Commun.} 
{\bf 108}, 147 (1998).

\bibitem{Li06a} Y.L. Li et al., {\sl Phys. Rev. B} {\bf 73} 184112 (2006).

\bibitem{Wang04} Y.U. Wang et al. {\sl Acta Mat.} {\bf 52} 81 (2004).

\bibitem{Semenovskaya97} S.Semenovskaya and A.G. Khachaturyan, {\sl Acta Mat.} {\bf 45} 4367 (1997).

\bibitem{Roytburd02} A. Roytburd and J. Slutsker, {\sl Acta Materialia} {\bf 50}, 1809 (2002).

\bibitem{Li06b} Y.L. Li et al., {\sl Appl. Phys. Lett.} {\bf 88} 072905 (2006).

\bibitem{Arimoto04} Y. Arimoto and H. Ishiwara, {\sl MRS Bulletin} {\bf 29}, 823 (2004).

\bibitem{Scott05} J.F. Scott, {\sl Nature} {\bf 4} 13 (2005).

\bibitem{Fu03} H.X. Fu et al, {\sl Phys. Rev. Let.} {\bf 91}, 257601 (2003);
I.I. Naumov et al., {\sl Nature} {\bf 432} 737 (2004).

\bibitem{Gutkin00} M.Y. Gutkin et al., {\sl J. Phys. Cond. Mat.} {\bf 12},
5391 (2000); A.G. Sherman and M.Y. Gutkin, {\sl Phys. Stat. Sol.} 
{\bf 184}, 485 (2001); M.Y. Gutkin et al., {\sl J. Phys. Cond. Mat.} 
{\bf 15} 3539 (2003). 

\bibitem{Khomskii06} D.I. Khomskii, cond-mat/0601696.

\bibitem{Lawes05} G. Lawes et al, {\sl Phys. Rev. Let.} {\bf 95}, 87205 (2005);
A.B. Harris and G. Lawes, cond-mat/0508617.

\bibitem{Mostovoy06} M. Mostovoy, {\sl Phys. Rev. Let.} {\bf 96}, 67601 (2006).

\bibitem{Zheng04} H. Zheng et al., {\sl Science} {\bf 303} 661 (2004).

\bibitem{Ganpule00} C.S. Ganpule et al., {\sl Appl. Phys. Lett.} {\bf 77} 
3275 (2000).

\bibitem{Krishnan02} A. Krishnan et al., {\sl Integ. Ferroelectrics} {\bf 43} 31 (2002).

\bibitem{Fousek03} J. Fousek and L.E. Cross {\sl Ferroelectrics} {\bf 293} 
43 (2003). 

\bibitem{Colla00} E.V. Colla et al., {\sl Phys. Rev. Let.} {\bf 85} 3033 
(2000); E.V. Colla et al., {\sl Phys. Rev. B} {\bf 63} 134107 (2001).

\bibitem{Grupp97} D.E. Grupp and A.M. Goldman, {\sl Science} {\bf 276},
382 (1997).

\bibitem{Coleman05} P. Coleman, cond-mat/0512463; S.E. Rowley et 
al, unpublished.

\bibitem{Scott06b} J.F. Scott et al., {\sl J. Phys. Cond. Mat.} {\bf 18} L205 (2006).

\bibitem{Sachdev99} S. Sachdev {\sl Quantum Phase Transitions} 
(Cambridge University Press, Cambridge, 1999).

\bibitem{Roussev03} R. Roussev and A.J. Millis, {\sl Phys. Rev. B} {\bf 67}, 
014105 (2003).


\end{thebibliography}
\end{document}